\documentclass[aps,pre,floatfix,twocolumn,footinbib]{revtex4-1}
\usepackage{graphicx}
\usepackage{amsmath}
\usepackage{appendix}
\usepackage{amssymb}

\newcommand{\be}{\begin{equation}} 
\newcommand{\ee}{\end{equation}} 
\newcommand{\bea}{\begin{eqnarray}} 
\newcommand{\eea}{\end{eqnarray}} 
\newcommand{\bc}{\begin{center}} 
\newcommand{\ec}{\end{center}}

\begin{document}

\title{Metanetworks of artificially evolved regulatory networks}

\author{Bur\c cin Danac{\i}}
\author{Ay\c se Erzan}

\address{Department of Physics Engineering,Istanbul Technical University, 
Maslak, Istanbul, Turkey}

\begin{abstract}

We study metanetworks arising in genotype and phenotype spaces, in the context of a model population of Boolean graphs evolved under selection for short dynamical attractors.  We define the adjacency matrix of a graph as its genotype, which gets mutated in the course of evolution, while its phenotype is its set of dynamical attractors. Metanetworks in the genotype and phenotype spaces are formed, respectively, by genetic proximity and by phenotypic similarity, the latter weighted by the sizes of the basins of attraction of the shared attractors.  We find that populations of evolved networks form giant clusters in genotype space, have Poissonian degree distributions but exhibit hierarchically organized $\kappa$-core decompositions. Nevertheless, at large scales, they form tree-like expander graphs. Random populations of Boolean graphs are typically so far removed from each other genetically that they cannot form a metanetwork. In phenotype space, the metanetworks of evolved populations are super robust both under the elimination of weak
  con\-nections and random removal of nodes.  



\end{abstract}
\date{\today}
\maketitle

\section{Introduction}

Robustness under mutations and ability for innovation give rise to evolvability of biological networks~\cite{Wagner1}. Robustness means the capacity of a population to explore a broad range of genetically accessible solutions to a particular evolutionary problem, without total loss of viability.  Innovation is the acquisition of new traits which enables the adaptation of the population to different circumstances.  In this paper we study the emergence, the topological properties and the robustness of metanetworks formed in phenotype as well as in genotype spaces in the course of the evolution of model gene regulatory networks (GRN).  We use populations  of Boolean graphs~\cite{Kauffman1992,Kauffman2004} evolved under selection for short dynamical attractors, with the assumption that only GRN with a predominance of point or period two attractors are viable~\cite{Danaci}.  

In a previous paper we found that the topological features  of the artificially evolved populations of Boolean graphs, described by the significance profiles of their motif statistics~\cite{Alon}, bear a close resemblance to those of gene regulatory networks of {\it  E. coli, S. cerevisiae} and {\it B. subtilis}.~\cite{Danaci} Other features, such as degree distributions, may vary  quite a bit from one population to another, since different  populations explore  different regions both in the genotype (adjacency matrix) and in the phenotype (attractor) space. The  diverse solutions to the same optimization process starting from different initial conditions, as well as the  slow, power-law relaxation to the  evolutionary steady state, indicates that the fitness landscape is a rugged one~\cite{Wright1932,Kauffman1987,Danaci}. This is a feature  encountered in spin glasses~\cite{Mezard,erzan1987}, a physical system which has a very large number of conflicting constraints.  

The different functions discharged by network motifs have also been investigated by Fran\c cois and Hakim~\cite{Francois2004design} by evolving  GRNs {\it in silico}. In a similar vein Burda {\it et al.}~\cite{burda2011motifs} have investigated which motifs emerge as dominant in GRNs which are either attracted to stable or multistable states  or perform switching functions, using Markov Chain Monte Carlo sampling  of small graphs.   

Ciliberti {\it et al.}~\cite{Ciliberti,ciliberti2} have studied the robustness and the capacity for  innovation of GRNs by studying the ``neutral network''~\cite{Nimwegen}  formed by individuals  within one mutational distance  from each other (i.e., genotypical neighbors)  displaying the {\it same phenotype}.  Capacity for innovation arises when this neutral network spans a large portion of the genotype space so that  many neighboring genotypes possess novel phenotypes.  To model this system, they choose what they   {\it define} as  ``viable''  networks, those which, starting from  unique, prescribed initial state eventually reach a single stationary state (attractor), which represents the phenotype.   However, Boolean graphs  may, in general,  have many attractors reached from different initial conditions. These different attractors can be shared by different sets of graphs.  

Allowing for more complex phenotypes (comprising more than one attractor) for each individual gives rise to the possibility of a complex weighted metanetwork in phenotype space. It is therefore worthwhile to study how the metanetwork of artificially evolved graphs within one mutational distance from each other,  spans  the metanetwork formed in phenotype space by  graphs with shared  phenotypical features. 

We have   adopted a selection bias for point or period two attractors in our artificial evolution algorithm. The motivation of this choice is as follows: in the context of differentiation into different cell types, where GRNs play a supreme role, it is necessary for the pattern of gene expression to be maintained stably once the cell differentiates. In the case of genetic switches, which may be components of more complex networks, the GRN may have more than one stable state and must be able to switch between them with some input from its environment~\cite{Cell_1}.
On the other hand various biological rhythms are regulated by gene clocks, which oscillate between different  patterns of gene expression.~\cite{Cell_2}  The simplest on-off oscillator is just a feedback system of two nodes  A and B with a repressing (A$\to$ B) and a positive (B$\to$A) interaction. The circadian clock has been modelled by Elowitz and Leibler~\cite{Elowitz2000} as a combination of repressor units producing a three-state cycle. We have focused on  point and period two attractors which seem to be the most dominant for small network units.

We define a metanetwork in the genotype space (MG) with nodes consisting of small Boolean graphs which differ from each other by one mutational step. We find that a finite population of randomly generated graphs do not form a metanetwork in genotype space, since the minimum``distance'' between them is seldom a single mutation. On the other hand, in almost all  of our evolved populations of Boolean graphs, more than half of each population belongs to a single connected component of the metanetwork.   Therefore, in the evolved population, the metanetwork in genotype space efficiently spans the phenotype space via single-mutational pathways and enabling innovation through new phenotypes.  This is an explicit example of the emergence of evolvability~\cite{Wagner1,Ciliberti,Nimwegen} in the course of evolution. 
 
\begin{figure}[!ht]
\centering
\includegraphics[width=7.9cm]{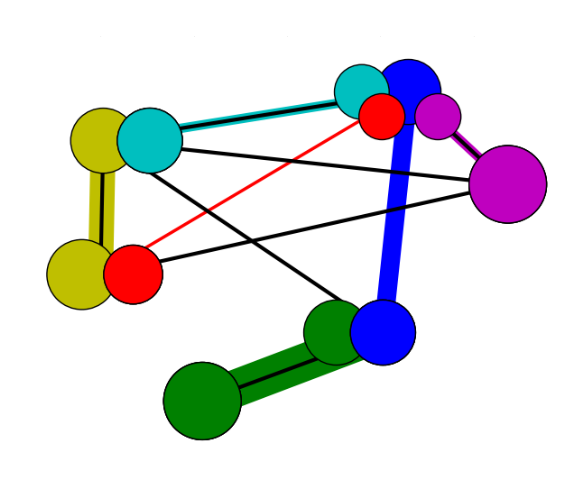}
\caption{(Color online) Illustration of a metanetwork in phenotype space. Each node (Boolean graph) is pictured as a cluster of balls. Balls of the same color correspond to  attractors that are shared between graphs associated with the nodes. Their radii represent the size of the basin of attraction of the attractor, for the dynamics at that node. The colored lines represent edges in the phenotype space and  black lines in the genotype space. The thickness of the colored edges are proportional to their weights ( see Eq.~\ref{eq:weights}).} 
\label{fig:pheno-met}
\end{figure}

We define a metanetwork in phenotype space (MPE) by requiring that  two Boolean graphs are connected by an edge, if they share at least one point attractor or an attractor with period two. The edges are weighted by the product of the sizes of their basins of attraction summed over the shared attractors.  The space of attractors shrinks to a small subspace in the course of evolution of our model, leading to strong edges and high phenotypic robustness. The evolved population consists of individuals which are capable of displaying a variety of behaviors given different sets of initial conditions and these behaviors overlap to a large extent between different individuals.

In Section II we define our model.  In Section III we present simulation results.  In Section IV  we discuss evolvability and robustness of the evolved metanetworks in genotype and phenotype spaces. Conclusions and a discussion are  provided in Section V.
\section{The Model}
A gene regulatory network is a collection of genes (nodes) which interact  (directed edges) with each other through the mediation of the proteins which they code.  These proteins (transcription factors) either activate or inhibit the transcription of the target gene whose transcription region they bind.  The GRN (or a module thereof, see e.g. the cell cycle module studied by  Li et al.~\cite{Li2004yeast}, Davidich and Bornholdt~\cite{Davidich2008boolean}) can be seen as an  automaton which, given a certain initial configuration of ``on" (1) and ``off" (0) genes, goes through a succession of states and  finally arrives at a steady state.  This steady state is termed an attractor in dynamical systems literature.~\cite{Ott2002,eckmann1981roads,strogatz1994nonlinear}  (see Appendix for definitions.)  

Dynamics on the GRN is modeled by defining variables $\tau_i,\; i=1\ldots N$ living on the nodes of the graph, and taking on the values of 1 or 0, corresponding to an active or a passive state of the node. The state of the system is  given by the vector $\mathbf{\tau}=(\tau_1, \ldots, \tau_N)$. The type of  interaction between pairs of nodes are  predefined and mutations only affect the topology of the graphs by changing elements $A_{ij}$ of the adjacency matrices.  The $A_{ij}=1$ ($A_{ji}=1$)  if $i$ and $j$  ($j$ and $i$) are connected in that order, and is zero otherwise.  
We assigned a random vector, a  Boolean ``key"  ${\mathbf  B}_j= (B_{1j}, \ldots,B_{ij}, \ldots, B_{Nj})$ to each $j$th node, with $B_{ij}$ determining the nature of the interaction with node $i$; $B_{ij}=1$, for a suppressing, or 0 for an activating, interaction respectively.  One set of $N$  keys are  randomly generated (with equal probabilities of zeros and ones) once and for all in the beginning of the simulations.  The matrix ${\mathbf B}$ is the same for all the graphs 
 and the keys ${\mathbf  B}_j$ have the convenient function of labeling the nodes $j=1,\ldots N$.   The synchronous updating rule is given by a majority rule, such that 
$\tau_{j} (t+1) = 1$ if  $ \sum_i^N  A_{ij} [\tau_i(t)\; {\rm XOR}\; B_{ij}]\ge N/2$ and is zero otherwise. 

In our model,  initial populations of random directed graphs are generated with a uniform edge density $p$. Each graph consists of $N=7$ nodes. The populations are evolved using  a genetic algorithm~\cite{Holland}, a standard procedure  for  solving optimization problems with an extremely large number of possibly conflicting degrees of freedom.  In the present case, the parameter that is optimized is  $a$, the attractor length averaged over all initial states.   We select for GRNs that have dominant attractors that are either fixed points or oscillators between two states.

The implementation of the genetic algorithm is as follows:
\begin{itemize}
\item At each step of the algorithm half of the graphs with the mean attractor length $a\le 2$ are chosen at random to be cloned.

 \item The chosen graphs are mutated by the standard edge-swapping approach. We randomly pick two independent pairs of connected nodes and switch either the in- or 
out-terminals of the edges.  This method preserves the in- and out-degrees of each node.  Four elements in the adjacency matrix of the graph change as a result. 

\item An equal number of graphs randomly chosen from the whole population are then removed.
\end{itemize}

The genetic algorithm was iterated until the average attractor length $a$ stabilized for 250 generations.  A steady state was achieved after 150 iterations on the average (over 16 populations of $10^3$ individuals each).   Measurements were taken both over a window of 100 steps within the steady state regime and at one given point in time and averaged over the different populations. See Table~\ref{tab:table} for the average values and standard deviations of the  attractor lengths and the  degree of the nodes. Further details of the simulations are explained in~\cite{Danaci}. The codes used in the simulations can be accessed at~\cite{kreveik}. 

It should be stressed that we do not have any intention or claims of modeling the process of evolution of GRNs {\it per se}; we  employ the  genetic algorithm as a generic tool for obtaining a steady state population with optimized values of a chosen parameter. Nevertheless, our definition of a mutation can be interpreted as a ``speeded up'' shorthand for the process  whereby mutations severe certain regulatory interactions while  possibly  establishing new ones~\cite{Wagner1998,Wagner2003}.   Switching the terminals rather than connecting the freed end to a randomly  chosen node makes sure that the connectivity of the graph is preserved.  Note that simply deleting some bonds at random would  just lead to a decrease of the density of the graph, which trivially leads to a shortening of the average attractor.~\cite{Kauffman2004,Aldana2003,aldana2003random}.

We define mutational distance ${\it d_{IJ}}$ between two graphs with adjacency matrices ${\mathbf A^I}$ and ${\mathbf A^J}$ as
\be 
d_{IJ}\equiv \frac{\sum_{k,l}^N |A^I_{kl}-A^J_{kl}|}{4}\label{eq:distance},
\ee 
where $I, J = 1,\ldots, {\cal N}$ index the individual graphs and ${\cal N}$ is the size of the graph population. The adjacency matrix of the metanetwork in genotype space (MG)  is therefore given by,
\be \tilde{A}_{IJ}\equiv 
\begin{cases} 1, & {\rm if } \;\;d_{IJ} \le 1 \\ 
0, & {\rm otherwise }  \end{cases}. 	
\ee 
%
\begin{figure}[!ht]
\includegraphics[width=7.9cm]{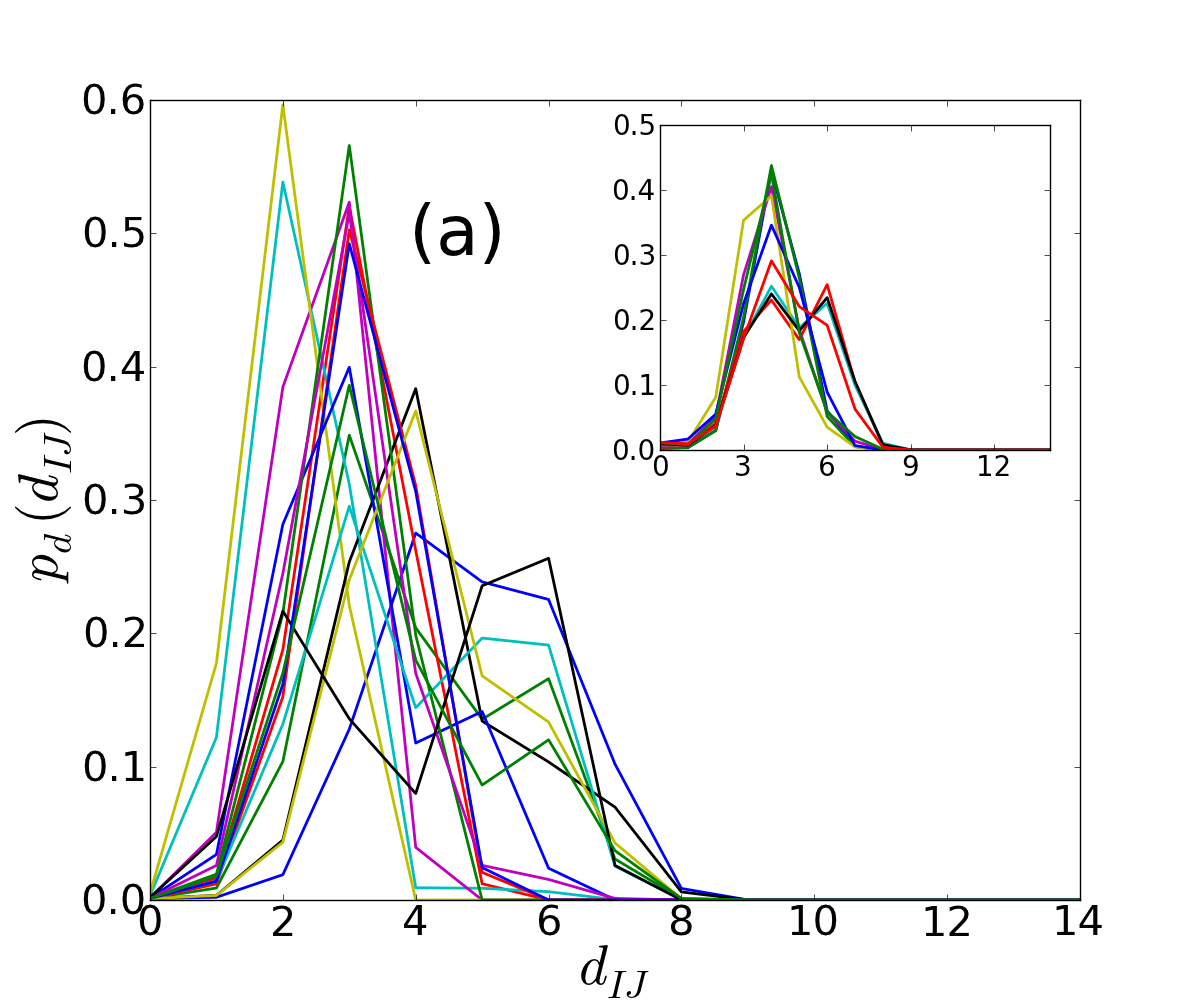}
\includegraphics[width=7.9cm]{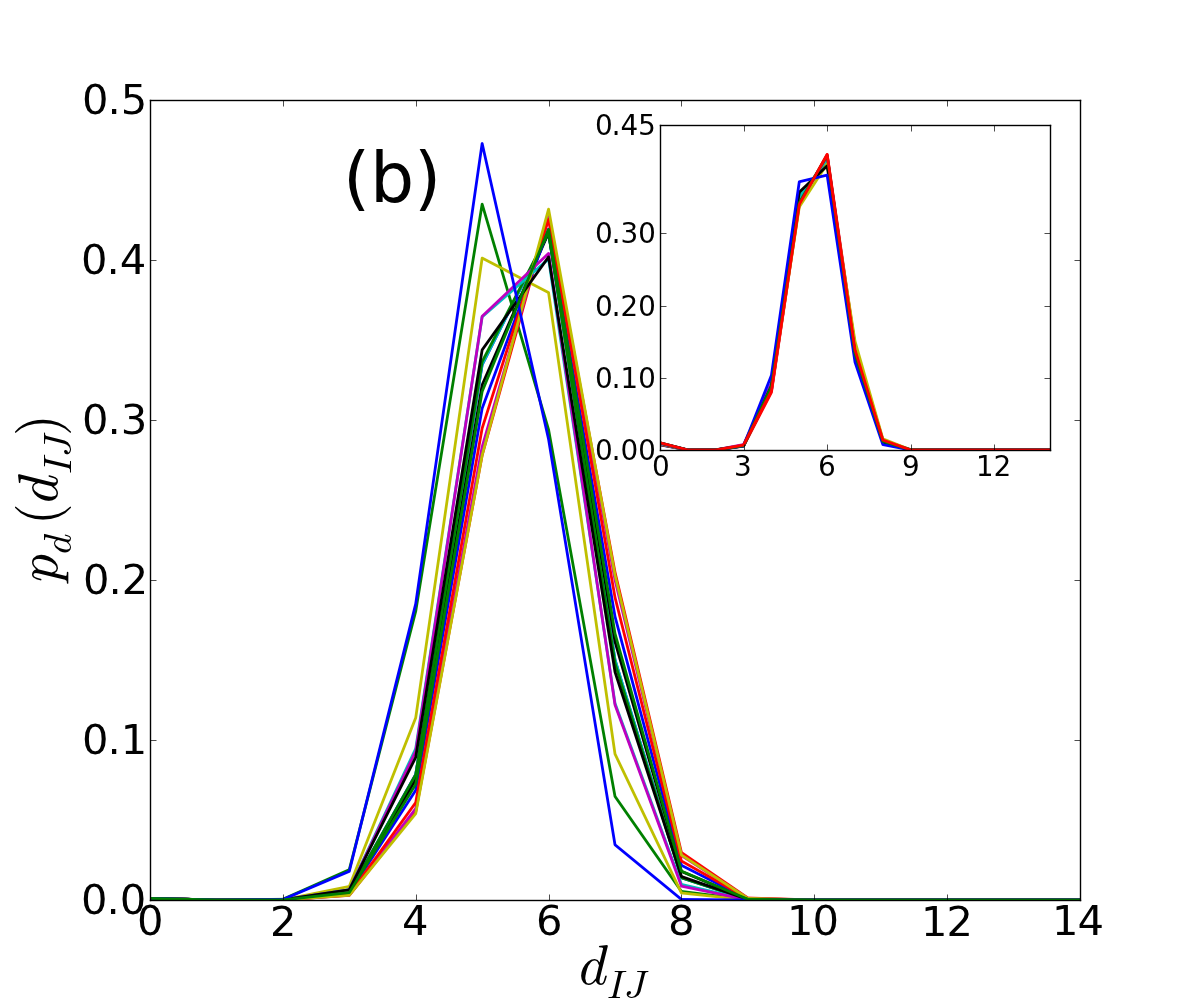}
\caption{(Color online) Distribution of the inter-graph distances $d_{IJ}$ in genotype space, for 16 independent sets of  (a)  evolved   and   (b) randomly generated populations of $10^3$ Boolean graphs each. The inset shows, for one population, the  pairwise distance distribution within the 10 largest clusters of graphs formed according to the criterion of sharing a given attractor.} 
\label{fig:neutral_distance}
\end{figure}
\begin{figure}[h]
\includegraphics[width=7.9cm]{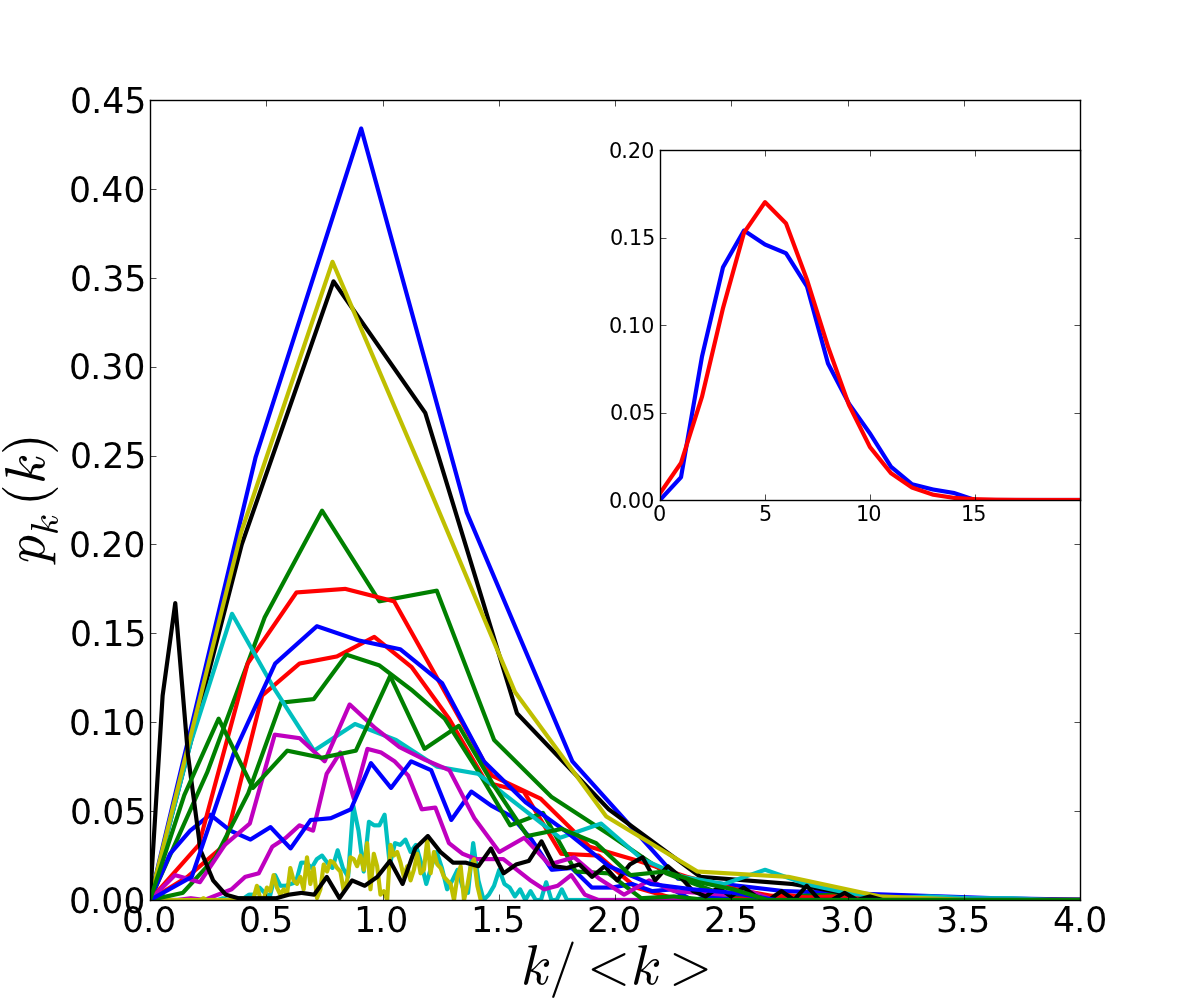}
\caption{(Color online) Degree distributions of metanetworks formed in genotype space by evolved populatios (MGE). The horizontal axis is normalized by the mean degree $\langle k\rangle$. The inset is the degree distribution of the MG of one of the evolved populations (blue) and  the Poisson distribution generated with the same mean value (red) appears only a little more peaked.} 
\label{fig:neutral_degrees}
\end{figure}

Next, we define metanetworks formed in phenotype space (MP).   Two graphs are connected by an edge if they have at least one viable  attractor in common (see Fig.~\ref{fig:pheno-met}). The weight of an edge is 
\be 
w_{IJ}\equiv \frac{\sum_{\alpha\in {\cal S}} \vert \Omega_{\alpha}^I\vert \vert \Omega_{\alpha}^J\vert }{{(2^N)}^2}\label{eq:weights},
\ee 
where  $\vert \Omega_\alpha^I\vert$ and $\vert \Omega_\alpha^J\vert$ are the sizes of the basin of attraction of the attractor $\alpha$ in the respective phase spaces of the Boolean graphs $I$ and $J$; ${\cal S}$ is the intersection of their sets of point or period two attractors. The weights are normalized  by the total phase space of the Boolean graphs $I,\,J$.

\section{Simulations and Results}

We have performed measurements on 16 independent populations of ${\cal N}=10^3$ randomly generated connected graphs with $N=7$ nodes each and an initial mean edge density $p=0.5$. The populations were evolved according to the genetic algorithm described in Section II. As mentioned in the Introduction, we found in~\cite{Danaci} that independently evolving populations find different solutions to optimizing the average attractor length and end up with different mean degrees (See~\cite{Danaci}, Fig.5).  The values of these mean degrees are given in Table~\ref{tab:table}. For comparison we have generated null sets consisting of an equal number of random Boolean graphs with the same edge density  and with the same Boolean functions assigned to their nodes as the evolved populations.

The random graphs sample the genotype space in a statistically uniform manner. We have deliberately taken 16 different random as well as evolved populations so that we are able to monitor the variability which comes from our finite sample size ($10^3$). The relevant parameters are provided in Table~\ref{tab:table}.

 The exponential growth of the size of the phase space (and therefore the possible number of attractors) with the graph size makes it prohibitively expensive to increase the graph size arbitrarily. Moreover, the topological features, i.e. significance profiles of the evolved graphs are found to be similar to those of the core graphs of biological networks with varying graph sizes~\cite{Danaci}.  This suggests that the main topological properties of the biological regulatory networks do not depend strongly 
on the graph size. 

The modular structure of gene regulatory networks~\cite{Newman2006modularity} with relatively small and denser modules strung together into larger units~\cite{Rodriguez2009basic,Francois2004design} has further encouraged us to keep the graph size small.  The “cell cycle module” of yeast  studied by Li et al.~\cite{Li2004yeast}, Davidich and Bornholdt~\cite{Davidich2008boolean} with $N=13$ and $N=9$  respectively, is a case in point.  The average degree is $\langle k \rangle=4.8$ and $4.9$,  yielding the  densities $p=\langle k \rangle/N= 0.37$ and $0.54$.   For comparison note that the densities for the whole (known) gene regulatory networks of {\it E. coli,  S. cerevisiae} and {\it B. subtilis}    are 0.0016, 0.00065, and 0.0016 respectively~\cite{Rodriguez2009basic} while the densities in the innermost $\kappa$-core are 0.28, 0.072, 0.089 respectively~\cite{Danaci}. 

We have taken  an initial edge density of  $p=0.5$, in order to allow for the selection of graphs with shorter attractors out of a population which potentially has much longer ones.  In creating the initial population of random graphs, we independently query whether each pair of nodes are to be connected or not with the initial probability $p$.  This results in an initial population of graphs with a binomially distributed number of edges and the average degree can drop (see Table~\ref{tab:table}) in the course of the iterations of the genetic algorithm.~\cite{Danaci} Larger graph densities, are known to lead to more “chaotic” behavior, i.e., longer periods in the case of finite graphs, with a critical threshold for the onset of chaoticity growing with the graph size.~\cite{Kauffman2004,Aldana2003,aldana2003random,Coppersmith2001,Drossel2005,Balcan2006}  

\begin{figure}[h]
\centering
\includegraphics[width=7.9cm]{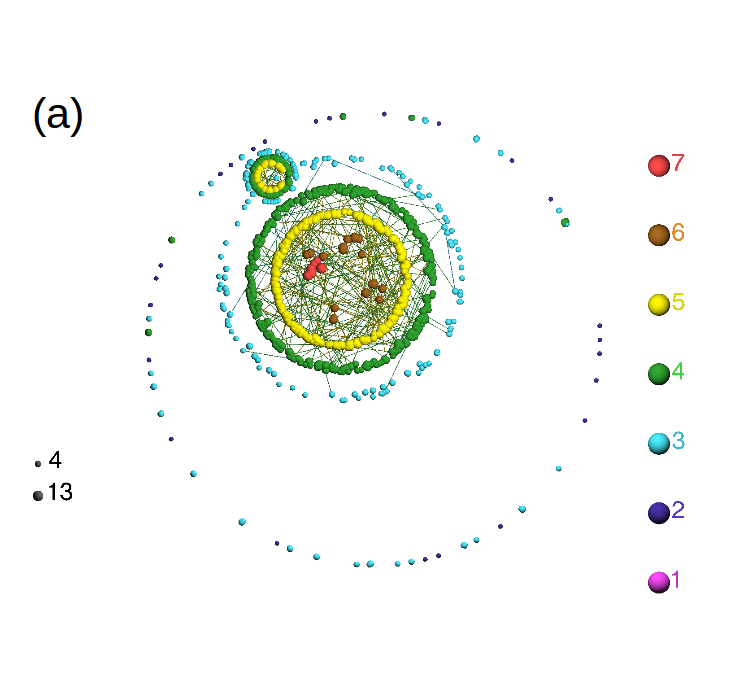}
\includegraphics[width=7.9cm]{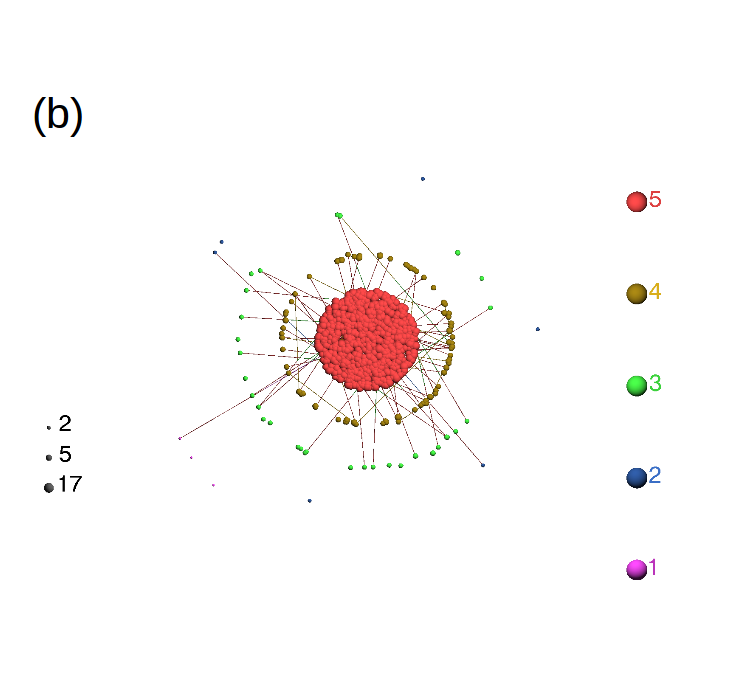}
\caption{(Color online) (a) $\kappa$-core decompositions of metanetwork formed by an evolved population in genotype space and (b) Erd\"os-Renyi network generated with the same edge density. The plots are obtained using Large Networks Visualization tool~\cite{Lanet}. The color code with the numbers on the side marks the $k$th shell within the network, the sizes of the balls (given on the left) grow with the degree of the node. The evolved population exhibits a more complex organization with 7 shells, or  levels of connectivity, v.s. only 5 for the random version.  The $\kappa$-core decompositions  for all 16 evolved populations are provided in Supplementary Material\cite{SM}. } 
\label{fig:$k$-cores}
\end{figure}
%
\begin{figure}[!ht]
\includegraphics[width=7.9cm]{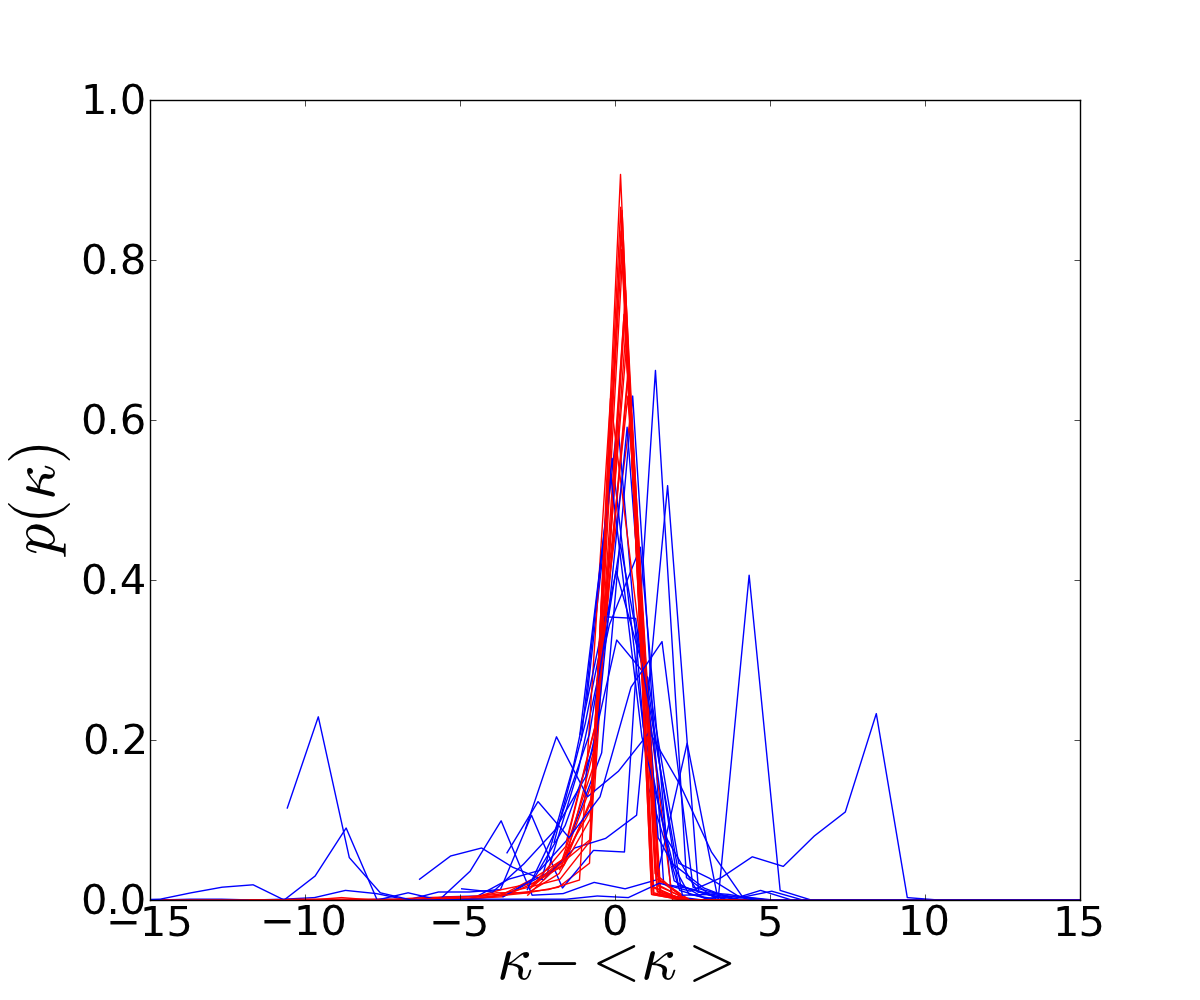}
\caption{(Color online) Distribution of the number of nodes found in the $\kappa$'th shell of the metanetworks in the genotype space for evolved populations (blue lines), and  surrogate Erd\"os-Renyi (E-R)  networks of the same size and with the same edge density (red lines), see text. } 
\label{fig:kshell_genotype}
\end{figure}

\subsection{Metanetworks in genotype space}

The distribution of the pairwise mutational distance $d_{IJ}$ for populations of the evolved and randomly generated graphs is given in Fig.~\ref{fig:neutral_distance}. The distance takes its maximum value when all the elements of the adjacency matrices of the two networks are different from each other, i.e. $|A^I_{kl}-A^J_{kl}|=1$ for all $k$ and $l$. So the maximum possible distance between two Boolean graphs is, for $N=7$,
\be 
d_{\rm max}\equiv \frac{\sum_{k, l}^7 1}{4}=12.25.
\ee 
For the evolved populations (Fig.~\ref{fig:neutral_distance}a), the distributions $p_d(d_{IJ})$ differ from set to set; however, most of them peak around $d_{IJ}=3$, which is about a quarter of $d_{\rm max}$. A few of the sets have distributions with a larger variance, exhibiting a second peak around $d_{IJ}=6$. We find that the MGEs of 13 out of the 16 populations exhibit a giant component of size $\ge 594$, {\it i.e.}, spanning  60\% or more of the whole population.

For the randomly generated populations, the mean pairwise distance $\langle d_{IJ}\rangle $ lies between 5 and 6, which is about half of $d_{\rm max}$ (Fig.~\ref{fig:neutral_distance}b) and $d_{\rm min}=2$, so that no metanetwork is formed in genotype space. The distributions $p_d(d_{IJ})$ are approximately symmetric around the mean values and similar for all the populations. 

The insets of Fig.~\ref{fig:neutral_distance} concentrate on one particular population. We consider clusters formed by Boolean graphs sharing a given attractor and plot pairwise distance distributions within the 10 largest such clusters. For the evolved population (Fig.~\ref{fig:neutral_distance}a), the distributions corresponding to different attractors are not the same, however they all have a peak at $d_{IJ}=4$. Some of the distributions exhibit yet another peak at $d_{IJ}=6$, which coincides with the mean inter-graph distance, $\langle d_{IJ}\rangle$, of the random graphs (Fig.~\ref{fig:neutral_distance}b). We conjecture that the two peaks of the evolved set correspond to the inter-cluster and intra-cluster distance distributions.
%
\begin{figure}[ht]
\includegraphics[width=7.9cm]{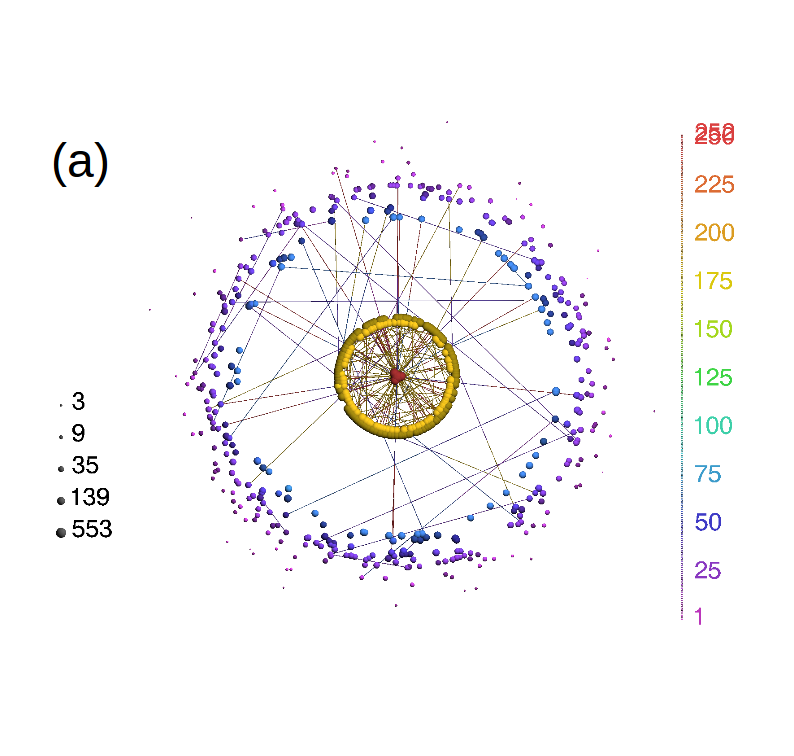}
\includegraphics[width=7.9cm]{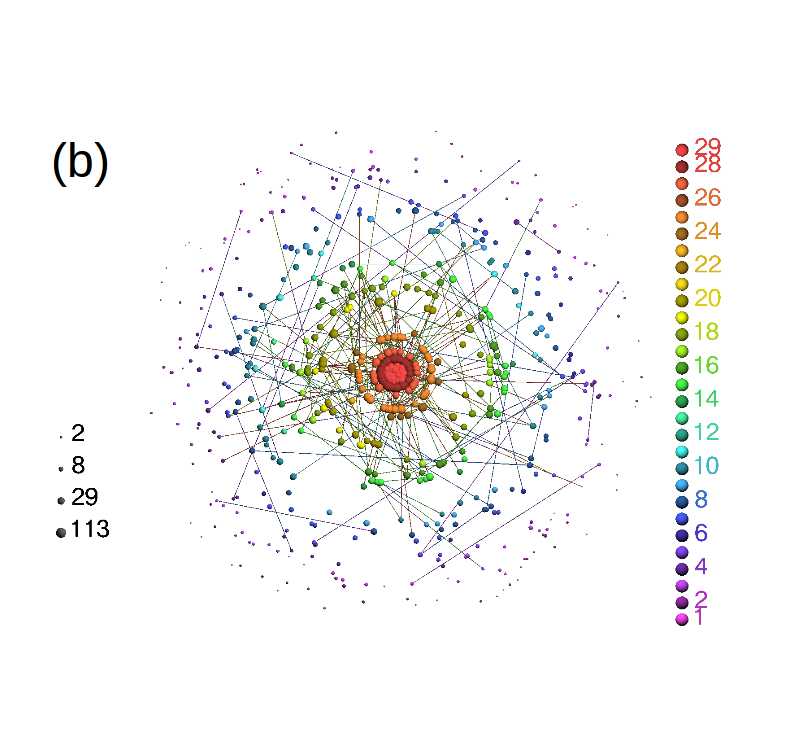}
\caption{(Color online) The $\kappa$-core decomposition~\cite{Lanet} of the phenotype metanetwork of an evolved (a) and randomly generated population (b).   The hierarchical organization of the metanetwork of the evolved graphs (MPE), going up to 252 shells, and degrees going up to $\sim 600$ is in stark contrast to the metanetwork of the randomly generated graphs (MPR) with only 29 shells and largest degree $\sim 150$. The $\kappa$-core decompositions  for all 16 evolved and random  populations are provided in Supplementary Material~\cite{SM}.} 
\label{fig:set2pheno}
\end{figure}
\begin{figure}[!ht]
\includegraphics[width=7.9cm]{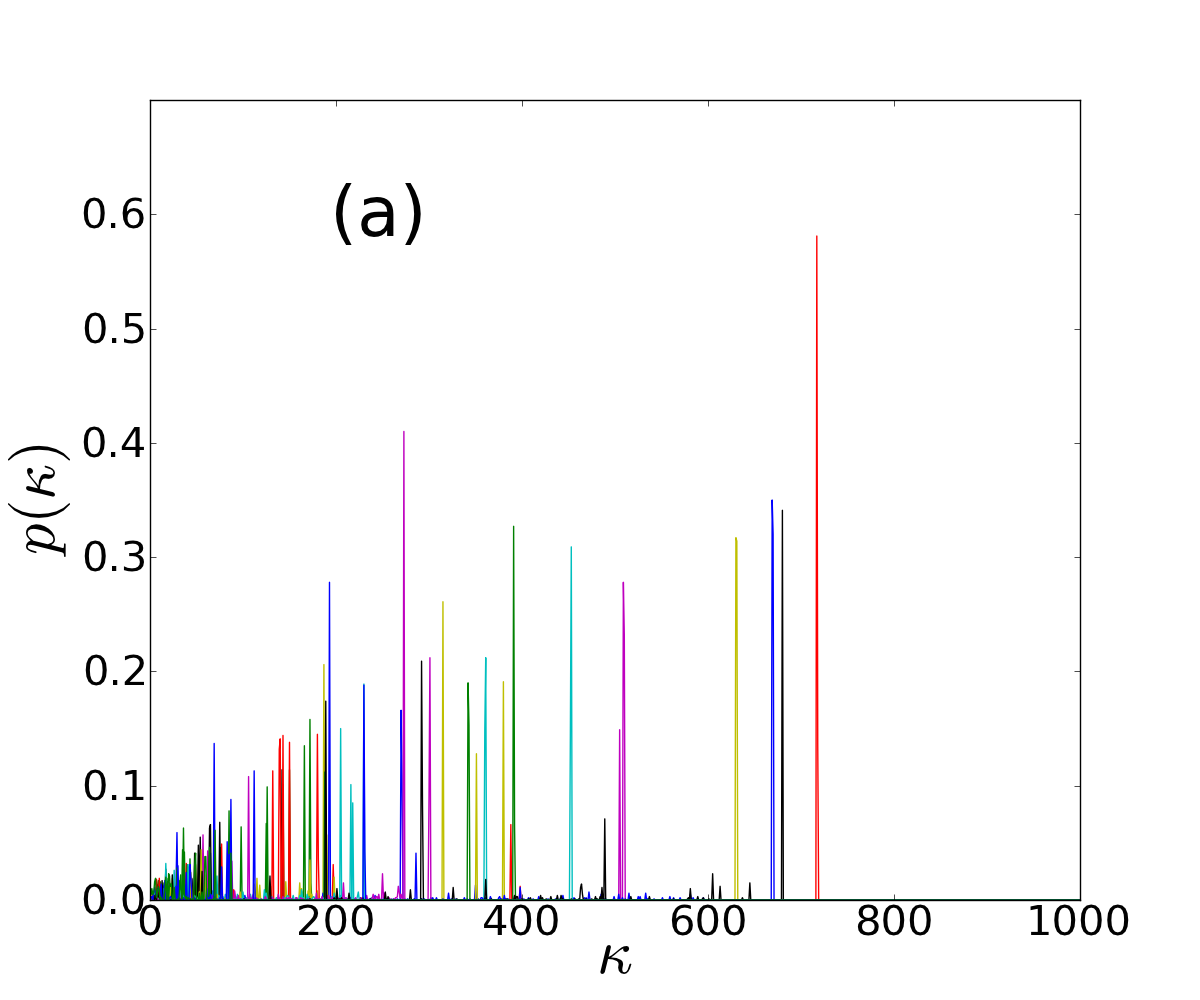}
\includegraphics[width=7.9cm]{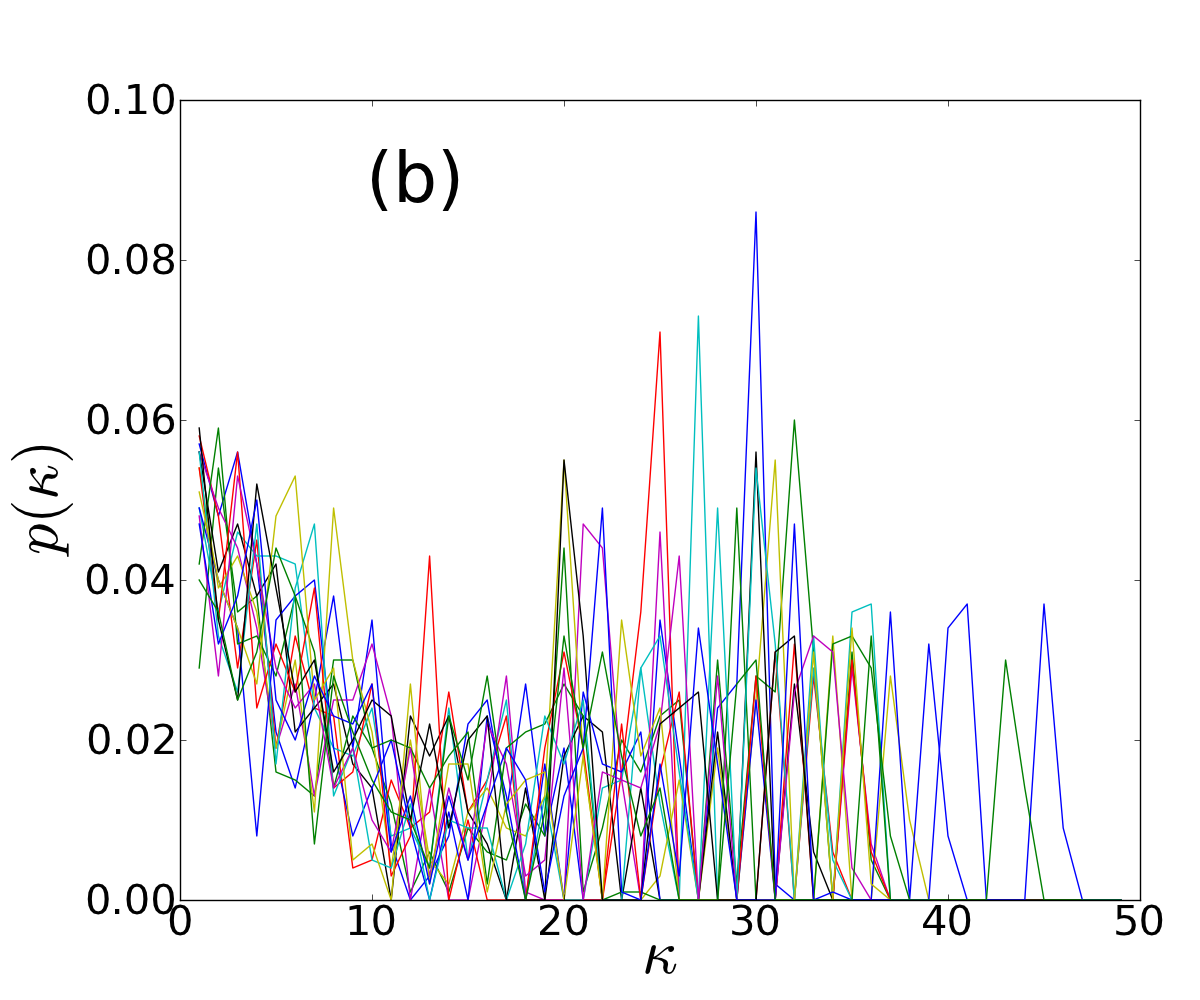}
\caption{(Color online) Superposed distribution of the number of nodes found in the $\kappa$'th shell of the metanetworks in the phenotype space by (a) evolved populations and (b) random populations. Distributions of the evolved sets are similar to the strength distribution depicted in Fig.~\ref{fig:strengthdist} and show a linearly increasing trend, while the random distributions are rather chaotic.  Note that the scales of the horizontal axes differ by two orders of magnitude.} 
\label{fig:kshell_phenotype}
\end{figure}
%
\begin{figure}[ht]
\includegraphics[width=7.9cm]{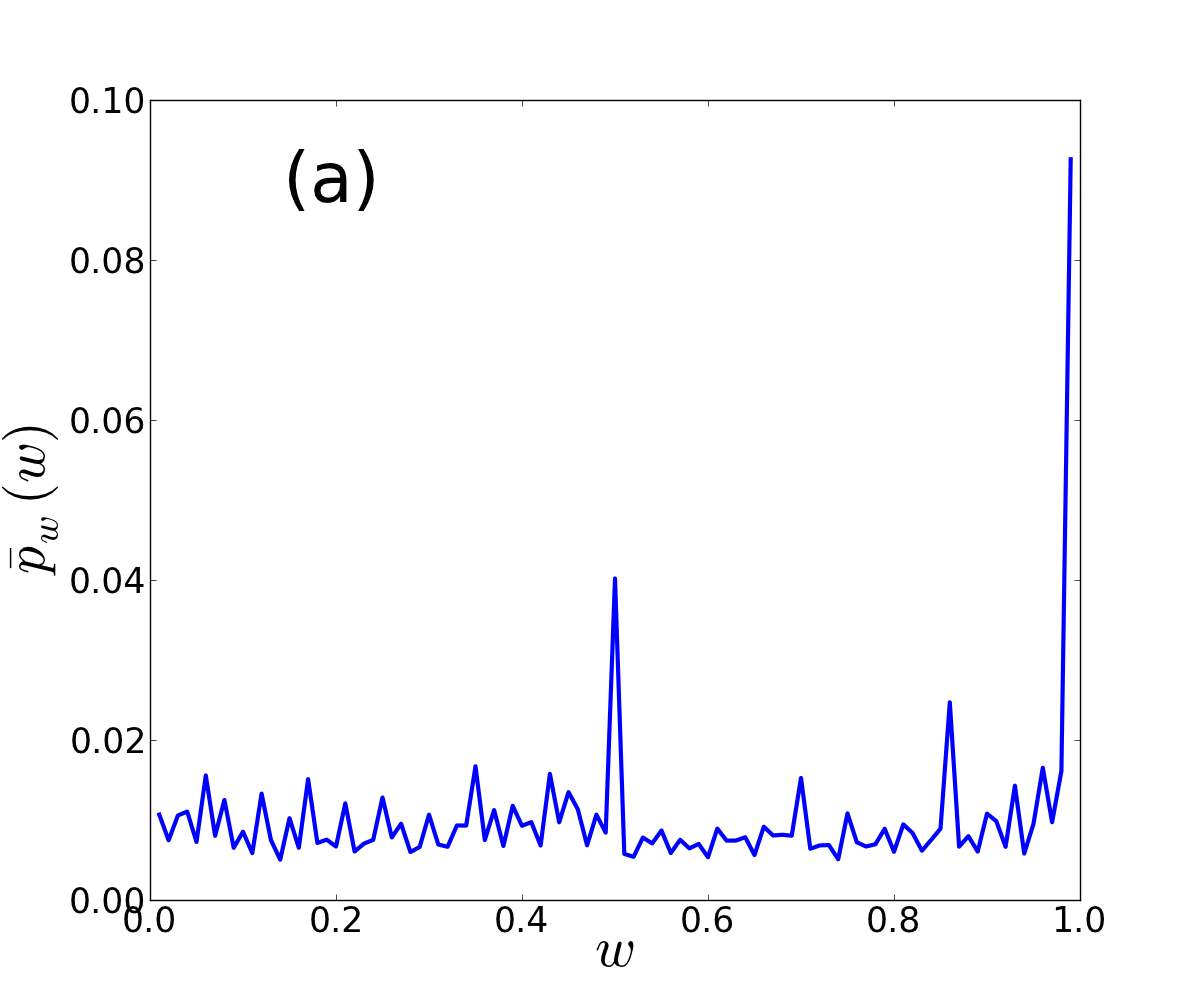}
\includegraphics[width=7.9cm]{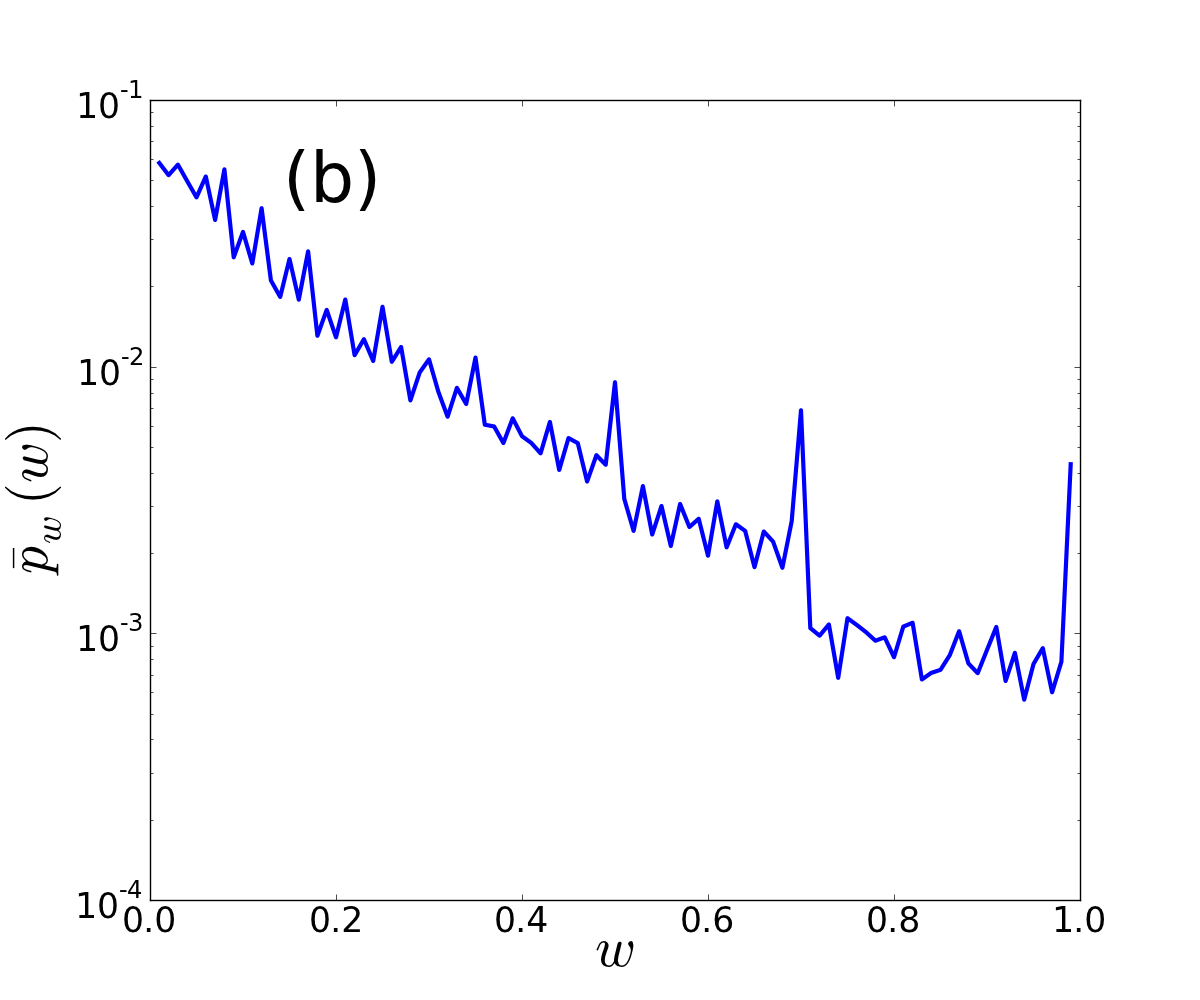}
\caption{The average weight distributions of the metanetworks formed in phenotype space of the evolved (a) and random (b) populations.} 
\label{fig:weightdist}
\end{figure}
%

The degree distributions of the MGE are shown in Fig.~\ref{fig:neutral_degrees}. Most of the distributions are approximately Poissonian, which suggests that these metanetworks might essentially be random. However, when we compare the $\kappa$-core decompositions of the metanetwork of an evolved set with that of an Erd\"os-Renyi network with the same edge density, we find that they can be quite different (Fig.~\ref{fig:$k$-cores}) and that the MGE displays a lot more structure.  See Supplementary Material~\cite{SM} for the $\kappa$-core decomposition of all the other evolved sets. 

In Fig.~\ref{fig:kshell_genotype}, we display the distribution of the nodes over the different $\kappa$-shells, for the 16 different evolved populations. Since  populations of randomly generated graphs with the same edge density do not form metanetworks in genotype space in general,  16 surrogate Erd\"os-Renyi (E-R)  networks of the same size (${\cal N}=10^3$) and with the same steady state edge density have been  generated, for comparison and their distributions are also presented in Fig.~\ref{fig:kshell_genotype} as  red lines.  Distributions of E-R networks are narrow and approximately symmetric around the mean shell number $\kappa$. Distributions of the metanetworks formed by the evolved populations are broader, not symmetric around the mean shell number and differ widely from each other due to the ruggedness of 
the fitness landscape as already  discussed above. The mean shell numbers and their standard deviations are given in Table~\ref{tab:table}.
\subsection{Metanetworks in Phenotype space}
The topology of the metanetwork in phenotype space shows striking differences between evolved and random sets. In this subsection we will investigate their connectivity properties.  In the next subsection we will explore their robustness under the  filtering of weak bonds as well as  the random removal of nodes.  

The metanetworks in phenotype space (MPE and MPR) are established by using the edge weights as defined in Eq.~\ref{eq:weights}. They both exhibit giant components.  In Figs.~\ref{fig:set2pheno} we display the visualization of the  $\kappa$-core decomposition of an evolved metanetwork in phenotype space. The corresponding "random" metanetwork is formed using the same rule, Eq.~\ref{eq:weights}, but on the population of random Boolean Networks with the same edge density as the evolved networks. In carrying out the  $\kappa$-core decomposition, we have considered all non-zero weights to be unity.  In Fig.~\ref{fig:kshell_phenotype}, we display the superposed distribution of the nodes over the different $\kappa$-shells. Note that the superposed  distribution of the MPEs show a linearly increasing trend.

In Fig.~\ref{fig:weightdist}, we present the weight distributions exhibited by the MPE and the MPR. The average weight distribution of the evolved sets (Fig.~\ref{fig:weightdist}a) is very flat with the prominent peaks at 1 and 0.5 carrying the imprint of the size distribution of the basins of attraction (Fig. 3.9 in ~\cite{danaci_thesis}), where, for both the evolved and random sets,  approximately 30 \% of the graphs have a dominant  basin of attraction occupying the whole phase space, and   another 10 \%  have basins of attraction at half this size. Those graphs with identical attractors occupying the whole of their phase space are joined by edges having  $w=1$, and this gives rise to extremely dominant hubs at the innermost $\kappa$-core of the MPE (Fig.~\ref{fig:set2pheno}a). In Fig.~\ref{fig:weightdist}b, the average weight distribution of the random sets shows an exponentially decaying trend with some trace of the peaks in panel (a) still surviving. The less pro\-noun!
 \-ced peaks lead to a loose conglomeration in the $\kappa$-core structure of the MPR (Fig.~\ref{fig:set2pheno}b).  

\begin{figure}[h]
\includegraphics[width=7cm]{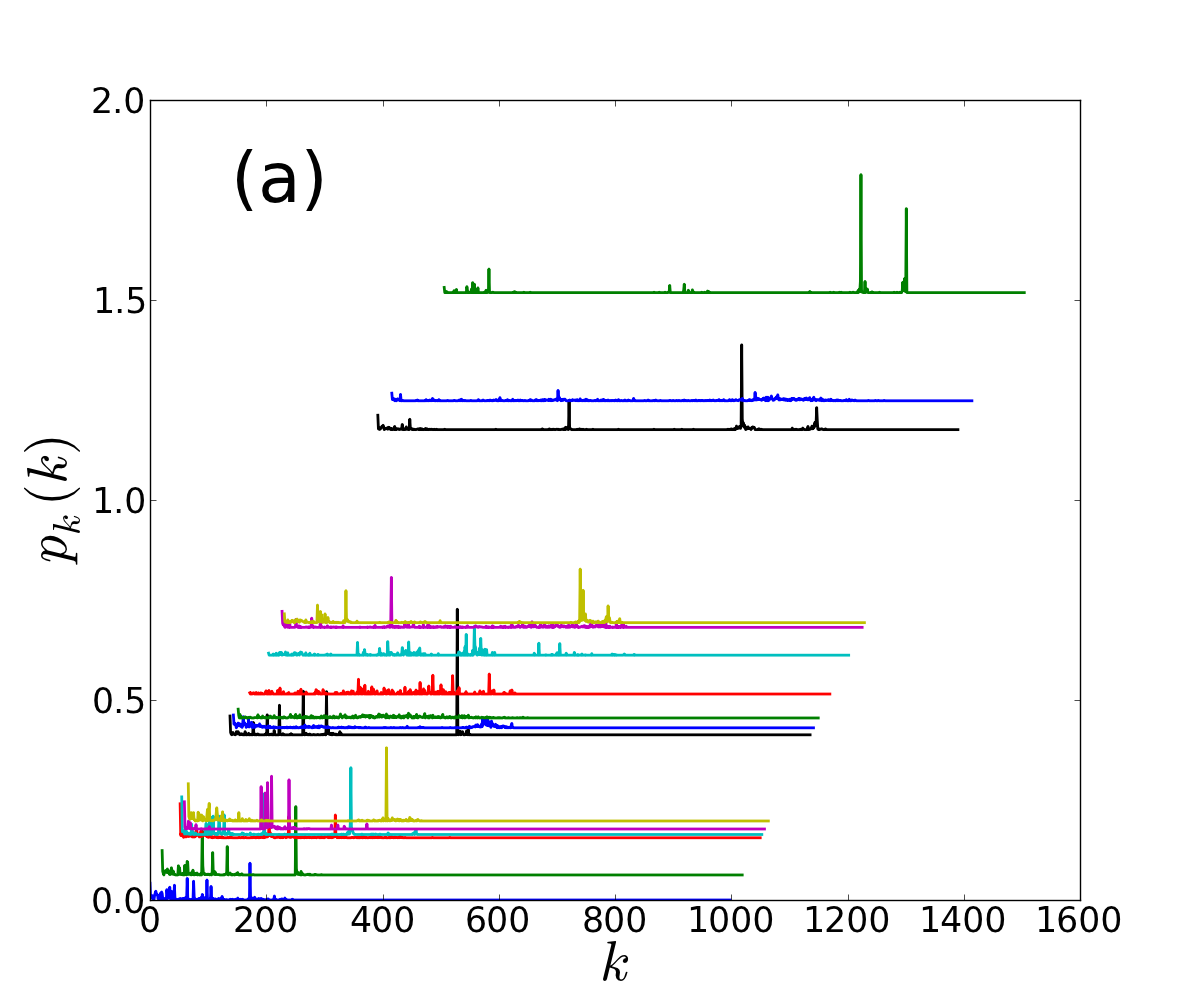}
\includegraphics[width=7.5cm]{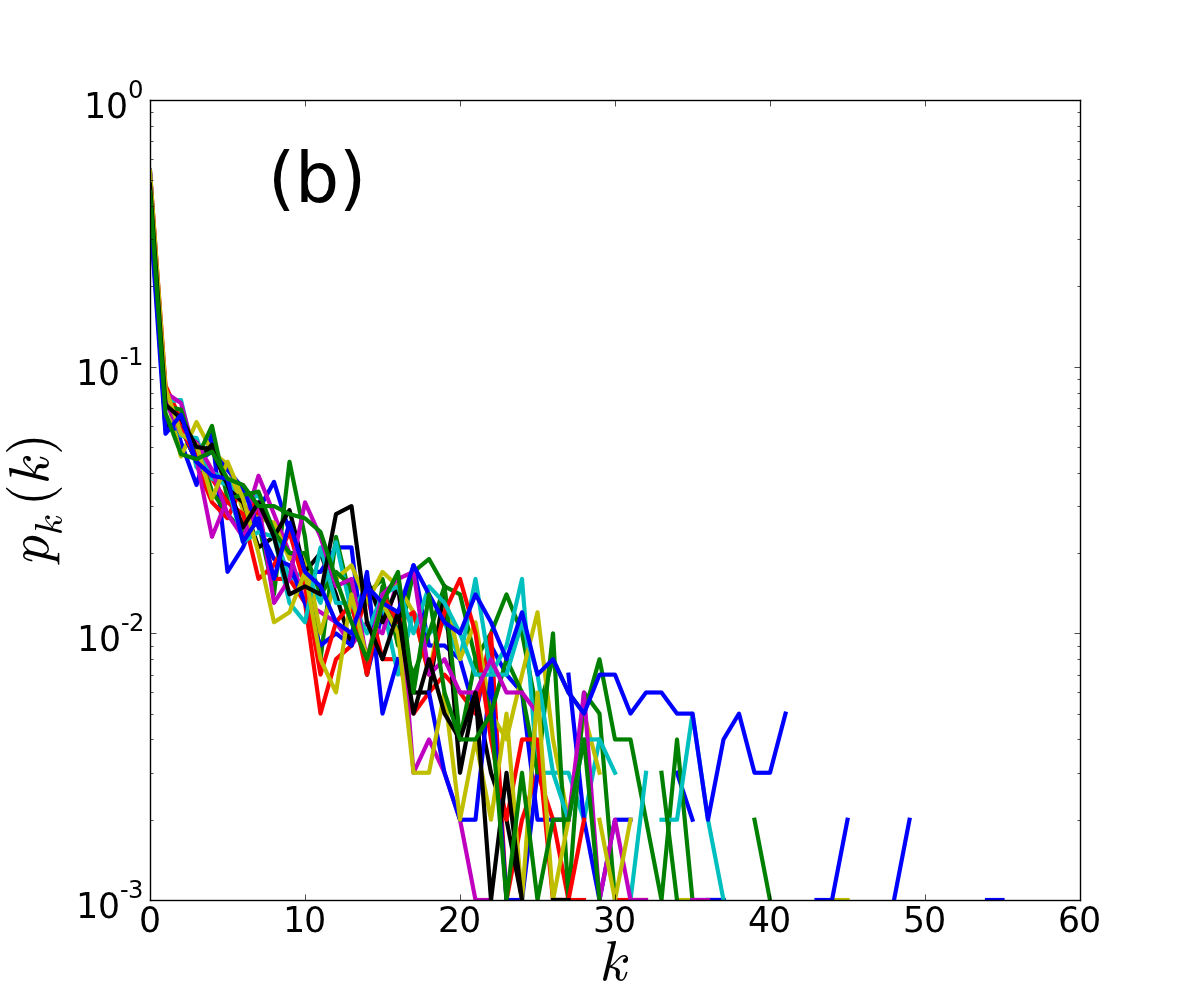}
\caption{ Superposed degree distributions of the metanetworks in phenotype space of 16 independent sets (a) of evolved graphs and (b) of randomly generated graphs, for $\theta=0.2$. For better visibility, The MPE degree distributions  have been superposed with an offset to the right on the $k$ axis  by $ \Delta_i = \langle k\rangle_i - \overline{k} $, where the $i$th set is ranked according to its mean degree $\langle k\rangle_i $ and  $\overline{k}=252 $ is the overall average degree with a standard deviation of  $\sigma_k= 144$.  The offset with respect to the  vertical axis is $0.003 \times\; \Delta_i$.  b) The degree distributions of the MPR for 16 independent sets of randomly generated graphs all display exponential tails on a semi-logarithmic plot and fall right on top of each other, with $\overline{k}=4.5 $ and $\sigma_k= 1.3$.} 
\label{fig:alldegdist}
\end{figure}

The degree distributions of the MPEs  are displayed in Fig.~\ref{fig:alldegdist}.  They are both quantitatively and  qualitatively different from those of the MPRs, and also from the degree distribution of the MGs, which are Poissonian.  In order to filter out the weakest bonds for better resolution,  we have omitted bonds with   $w_{IJ} <  0.2$  in computing the degree distributions.  The degree distributions of the MPEs vary strongly from set to set. The majority of the individual distributions exhibit two different components, the first being an exponential distribution confined to small $k$, and the second a series of outlier peaks whose height increases with $k$. The degree distribution of the random sets, however, decay exponentially.

We now turn to ``strength''  distribution of the  metanetworks in phenotype space. The ``strength'' of a node is defined~\cite{barrat2004architecture}  as the  total weight of edges impinging on the node, 
\be W_I=\sum_J w_{IJ}\;\;\;.
\label{eq:strength}
\ee  

\begin{figure}[h]
\includegraphics[width=7.5cm]{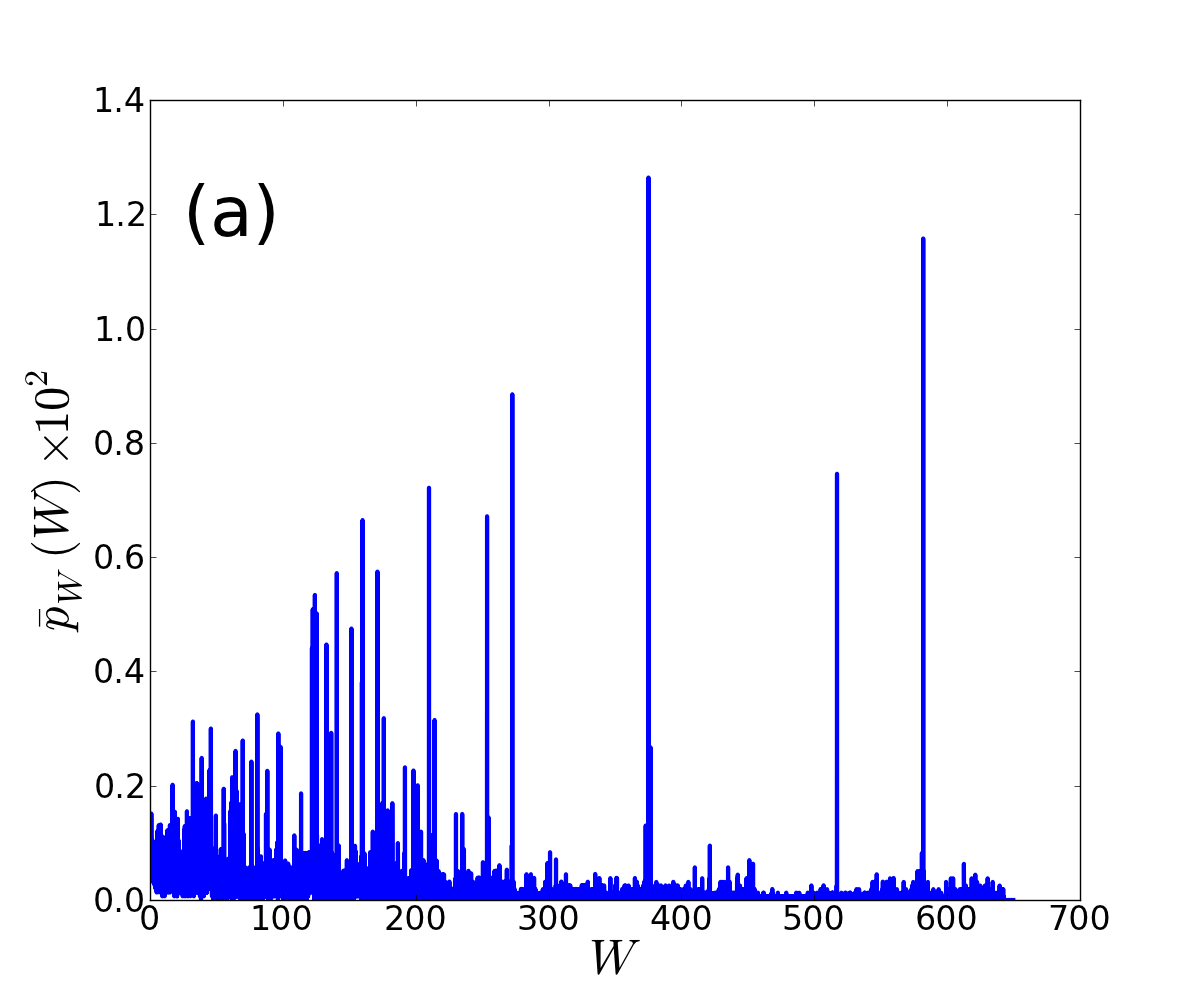}
\includegraphics[width=7.5cm]{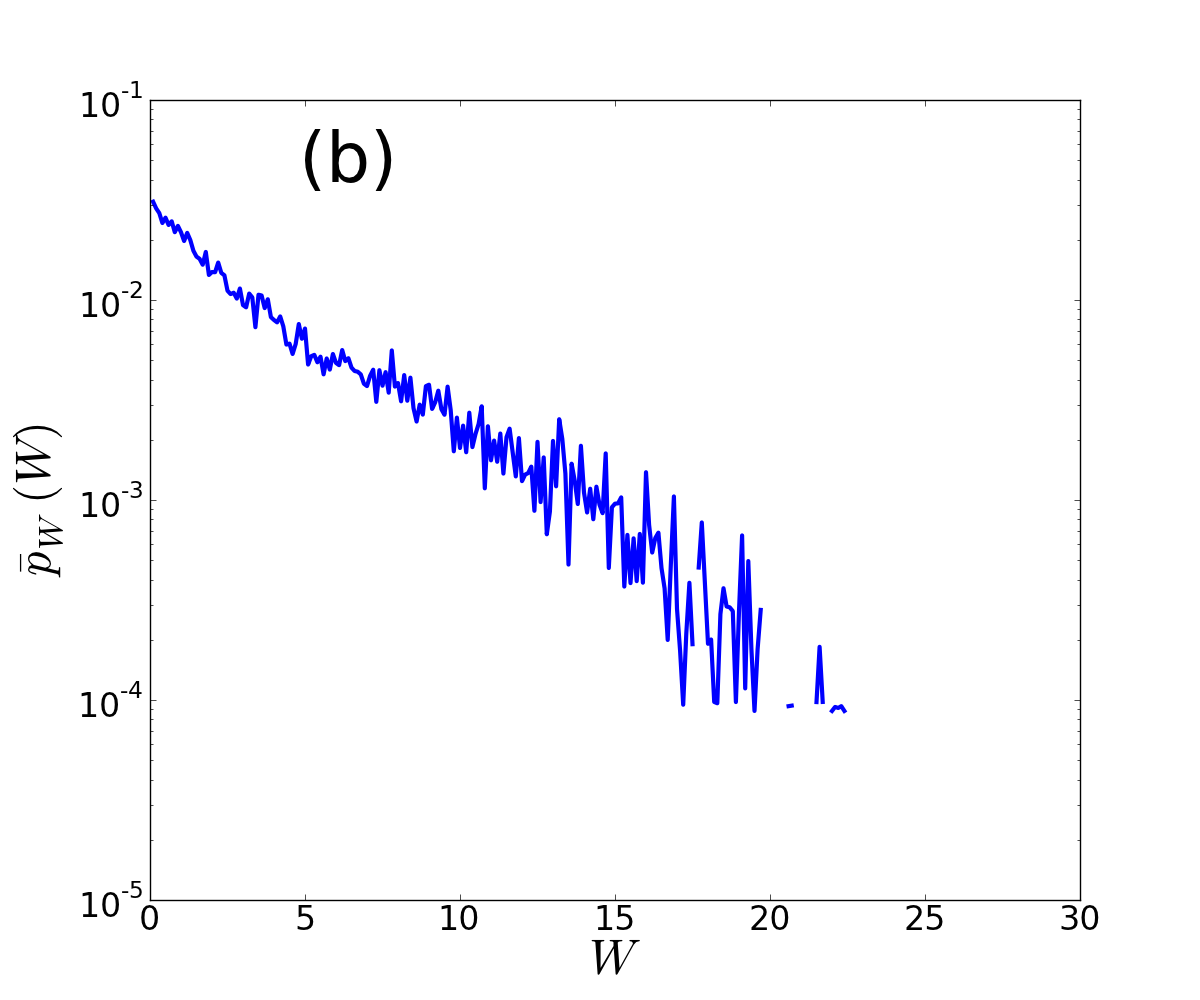}
\caption{The average distribution of the node strengths $W$  for the metanetworks in phenotype space of evolved graphs (a), and randomly generated graphs (b) (bin size = 0.1). Note that the vertical axis in (b) is 
logarithmic. The strength distributions for the evolved and the random  sets bear a very close resemblance to their $\kappa$-shell distributions (Fig.~\ref{fig:kshell_phenotype}). The evolved sets show a great deal of variation in the  $W$-distributions, as they do for the degree distributions.  The distributions for random sets fall right on top of each other and obey a Poisson distribution falling off exponentially; the horizontal scales differ by two orders of magnitude.} 
\label{fig:strengthdist}
\end{figure}

The strength distribution $p_W$  (Fig. \ref{fig:strengthdist}) is qualitatively very  similar to the distribution of the nodes over the $\kappa$-shells as depicted in Fig.~\ref{fig:kshell_phenotype}, as well as the degree distribution~\cite{barrat2004architecture}, see Fig.~\ref{fig:alldegdist}. 
In fact, it follows from Eq.~(\ref{eq:strength}) that, 
\be 
W_I = k_I \langle w\rangle_I\;\;.
\label{eq:W_dist}
\ee 
If the weight distribution $p_w$ is independent of the degree $k$, as suggested by  Fig.~\ref{fig:weightdist}a, then Eq.~(\ref{eq:W_dist}) simplifies to  $W_I =  \langle w\rangle k_I$.   We have checked that the correlation between the degree and the average weight,
\be c=\frac{\overline{\langle w\rangle_Ik_I}-\overline{\langle w\rangle}_I\overline{k}_I}{\overline{\langle w\rangle}_I\overline{k}_I}\;\;,
\label{eq:corelation}
\ee 
 is indeed small.  Here the overbar signifies an average over the nodes $I$, as well as over the independent populations. We find   $\langle c\rangle =  -0.003 $ and  standard deviation $\sigma_c =0.1$ over the sixteen sets.

%
%

\begin{figure}[h]
\includegraphics[width=7.5cm]{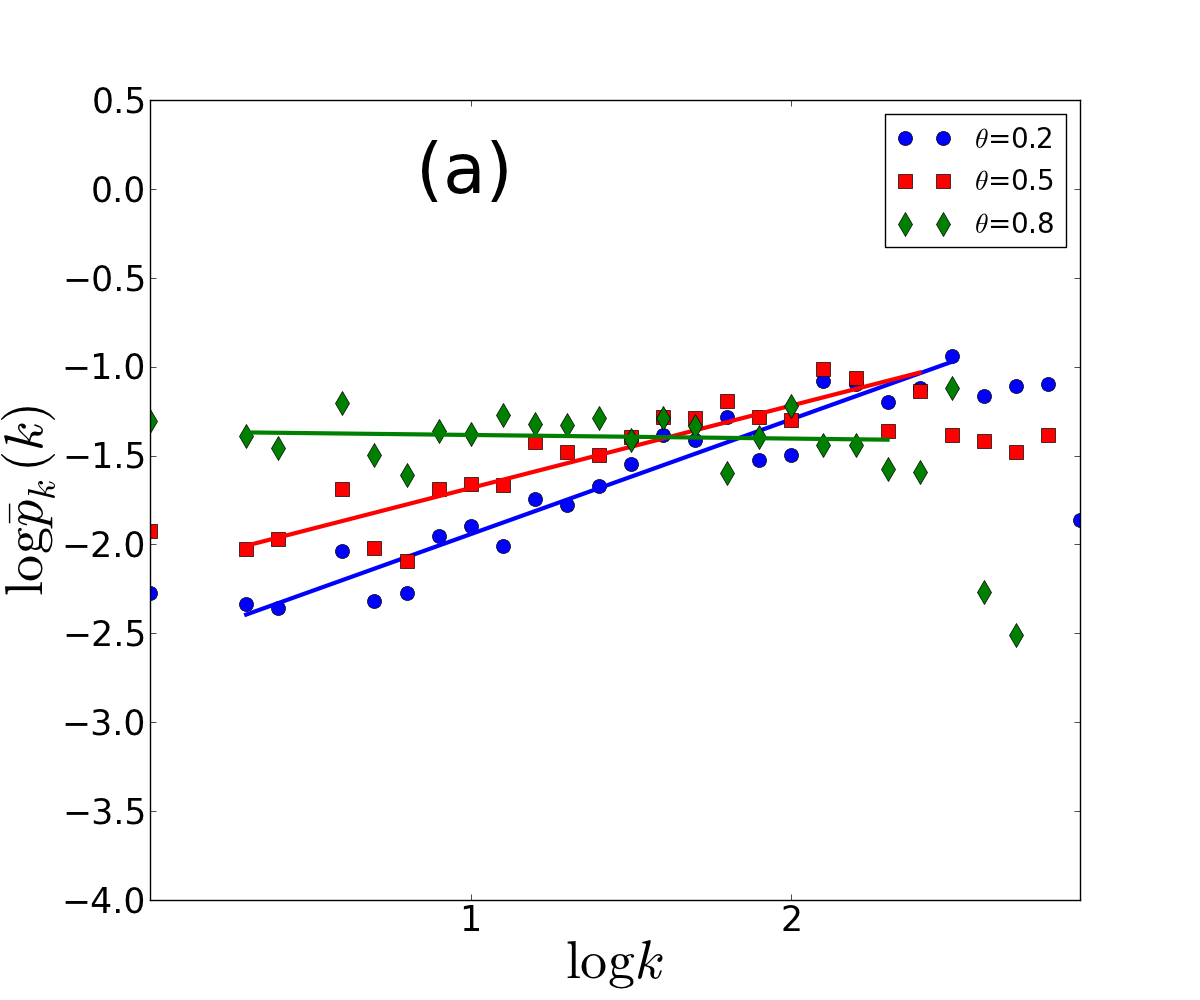}
\includegraphics[width=7.5cm]{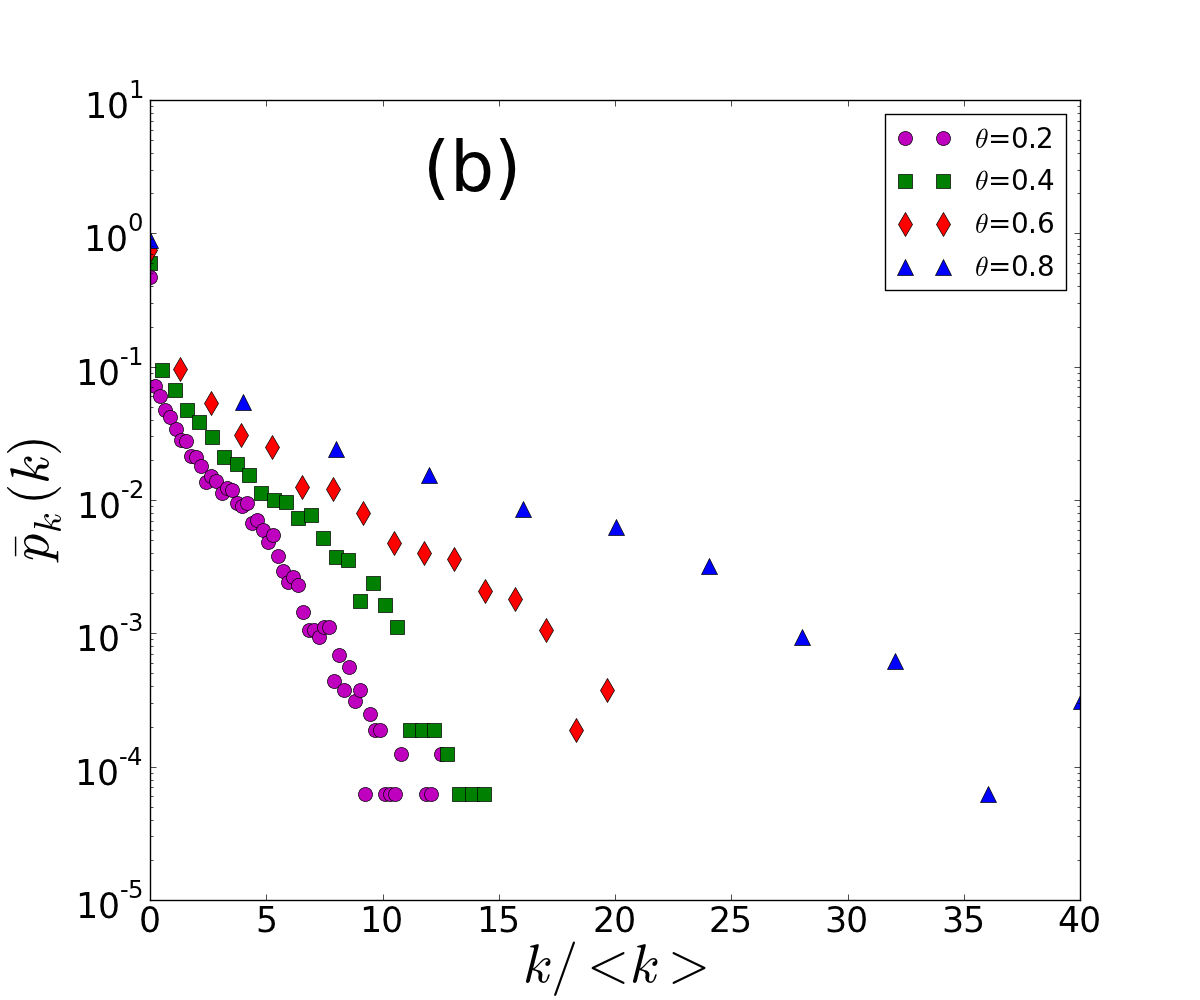}
\caption{(Color online) a) Log-binned degree distributions for the combined degrees of all 16 MPEs, with linear least square fits to three different $\theta$ values 0.2, 0.5 and 0.8, with $\gamma=-0.65\pm 0.04,\; -0.46\pm 0.05, \;0.02 \pm 0.05$, respectively.    b) Degree distribution of the MPR for 16 different independently generated random populations of Boolean graphs, for four different values of the weight threshold $\theta$ are displayed in this semilogarithmic plot. The horizontal axis has been normalized by $\langle k \rangle$. The distributions are exponential.} 
\label{fig:alldegdist_allthetas}
\end{figure}
\begin{figure}[ht]
\includegraphics[width=7.5cm]{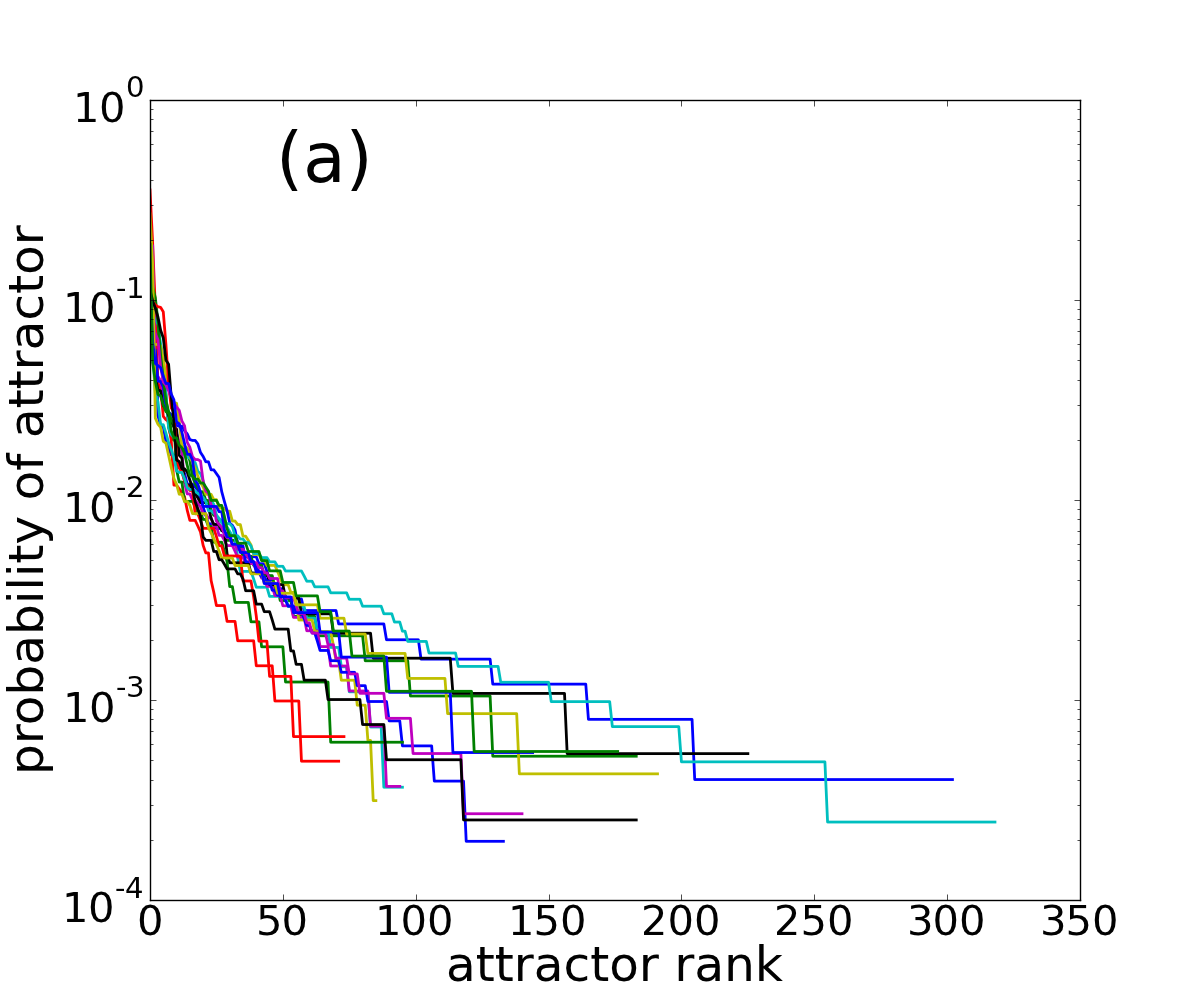}
\includegraphics[width=7.5cm]{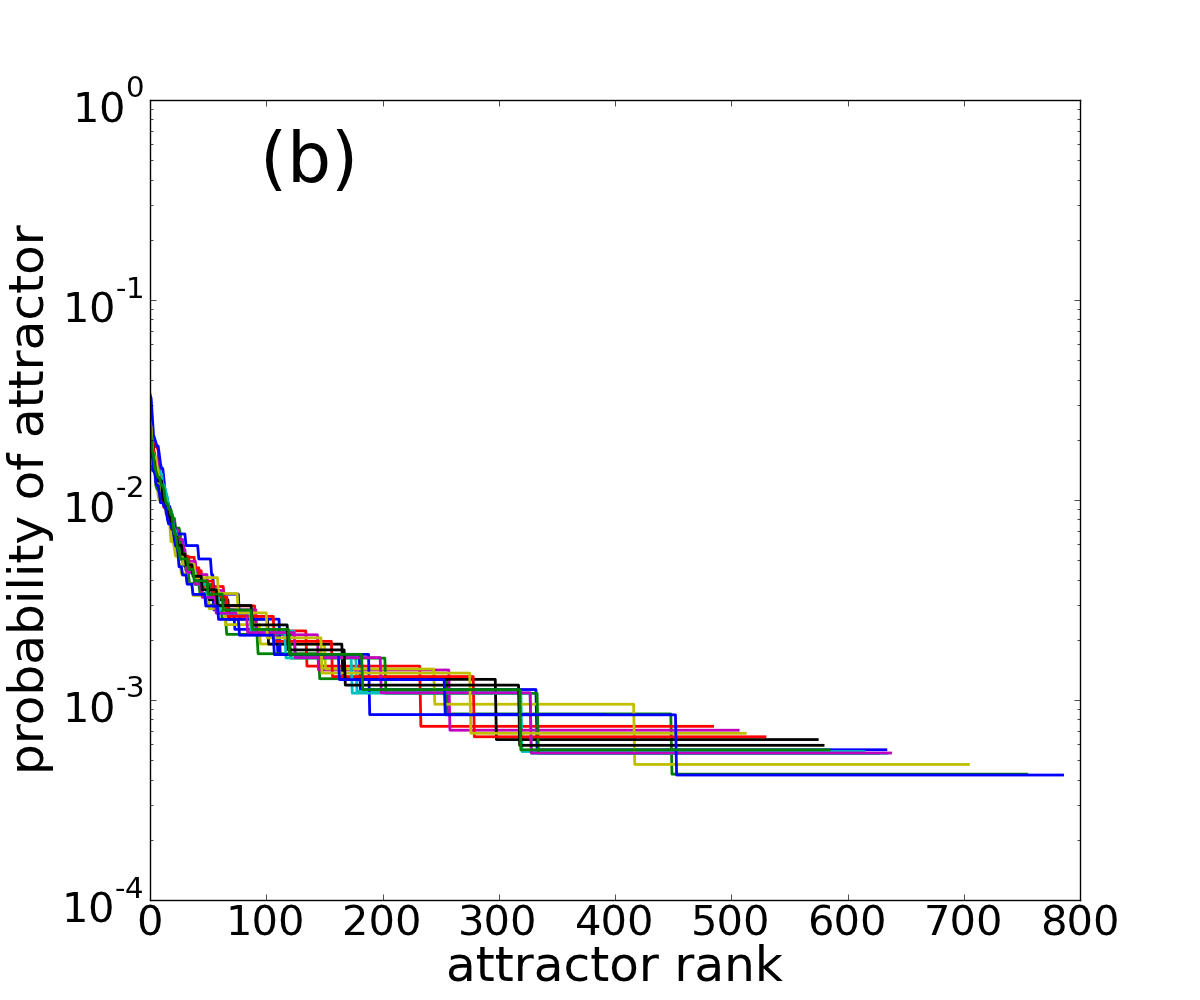}
\caption{(Color online) The  probabilities of occurrence  of point and period two attractors in (a) the evolved populations and (b) the randomly generated populations, ranked in order of their incidence.  The horizontal axes differ in their scales.} 
\label{fig:attractor_freqs}
\end{figure}

\section{Evolution of evolvability and robustness} 

\subsection{Convergence to regions of high connectivity} 

In order to understand the behavior of the metanetworks as we prune the edges by raising the weight threshold, as we will do in the next subsection, it is useful to first to see  how the degree  distribution changes as we change the weight  threshold $\theta$, below which a bond will be considered severed.  A bond between the pair of graphs $(I,J)$ is present  if $w_{IJ} \ge \theta$, and is otherwise set to zero. We can use this threshold as a ``filter'' for probing the structure of the MP.

In Fig.\ref{fig:alldegdist_allthetas}a, we present the combined, log-binned  degree distributions of all the 16 MPEs  for different values of $\theta$. Indeed, we find that for these evolved populations, the degree distribution of the MPE has an incipient power law form, $p_k \sim k^{-\gamma}$, albeit over a relatively small interval. The novelty here is that for $\theta < 0.8$, one finds $\gamma < 0$, i.e., the distribution {\it increases} as a function of the degree.  For the MPR, combining all the 16  random populations and performing a linear binning (bin size = 0.1), we see in Fig.\ref{fig:alldegdist_allthetas}b that the degree distribution is exponential.

Examples of networks with increasing rather than decaying power law degree distribution have been  observed by Barrat et al. \cite{barrat2004architecture} for transport networks, with $\gamma=-1$. We observe the same phenomenon for the combined $\kappa$-shell distribution, Fig.~\ref{fig:kshell_phenotype}a, and the combined strength distribution, Fig.~\ref{fig:strengthdist}a, where, for $\theta=0$, we have effectively found a {\it linear increase} in the shell population with the shell number and  the same linearly increasing  trend in the probability $p_W$ to encounter nodes with  strength $W$.   

The relatively  greater probability to find nodes with high connectivity is precisely what one means by a population to concentrate in regions with a high density of edges~\cite{Nimwegen} in genotype space; in our model, we see that this phenomenon holds also in phenotype space.  What is more, we see that  as $\theta$ grows, $\gamma$ becomes smaller in absolute value and  the degree distribution flattens out. Thus, for sufficiently stringent conditions for edges to form in phenotype space, one ends up with an almost  uniform degree distribution. For  $\theta \ge 0.8$ we see it becomes  very slightly positive.  

It is worthwhile to ask how the much stronger bonding between graphs in the evolved populations arise. 
The attractors shared by the Boolean graphs in the evolved populations have a narrower frequency distribution than those of the randomly generated populations. In Fig.~\ref{fig:attractor_freqs}a, we display, for 16 independently evolved populations, the normalized incidence, or probability, of different attractors v.s. their rank. In Fig.~\ref{fig:attractor_freqs}b, the same distribution for randomly generated graphs is displayed for comparison. The convergence to a small set of shared attractors (phenotypes) causes more and stronger bonds to form between the Boolean graphs in the evolved populations. 

%
\subsection{Robustness of the evolved and random metanetworks in genotype and phenotype space}

Kimura~\cite{Kimura} has introduced the concept  of neutral evolution, where an evolving  population is represented by different nodes in genotype space, populated by different numbers of individuals.  Pairs of nodes are connected by an edge only in case they differ from each other by just a single mutation.  The  connected set of relatively high fitness genotypes is termed the neutral network.  The robustness of a population subject to mutations can be thought of  as the average probability that an individual continues to reside on the neutral network after suffering a random mutation.   Nimwegen {\it et al.}~\cite{Nimwegen} show, for uncorrelated networks,  that this probability can be expressed in terms of the average degree of the nodes of the neutral network, weighted by the population  of each node.   They propose this average degree as the measure of robustness.  

In network theory there is another phenomenon which is termed robustness, which comes under the topic of percolation on networks~\cite{networkpercolation,Dorogovtsev-CCNW}.  Percolation on networks has been a central issue since the seminal paper by Albert and Barabasi~\cite{Albert} where they showed that scale free networks were resilient to random failure of nodes (say on a power or communications grid) but vulnerable to malicious attack, whereas random (E-R) graphs exhibited a percolation threshold at a finite fraction of removed nodes.   A network which retains  a finite connectivity until  the ratio of removed nodes is taken to unity (while the size of the system is taken to infinity), is called robust. 

In this subsection we first examine the metanetworks in genotype space (MGE) and in phenotype space (MP) and their response to the  random removal of nodes (note that the random Boolean graphs do not form a metanetwork in genotype space). 
Next we  study how the metanetworks in phenotype space, which span essentially the whole population for zero threshold ($\theta=0$), shrink as the value of $\theta$ is increased, leading to the removal of weak edges. For practical purposes, we define the giant component of the MG, as the set of nodes found in the largest component spanning more than 50\% of the nodes.  

\begin{figure}[h]
\includegraphics[width=7cm]{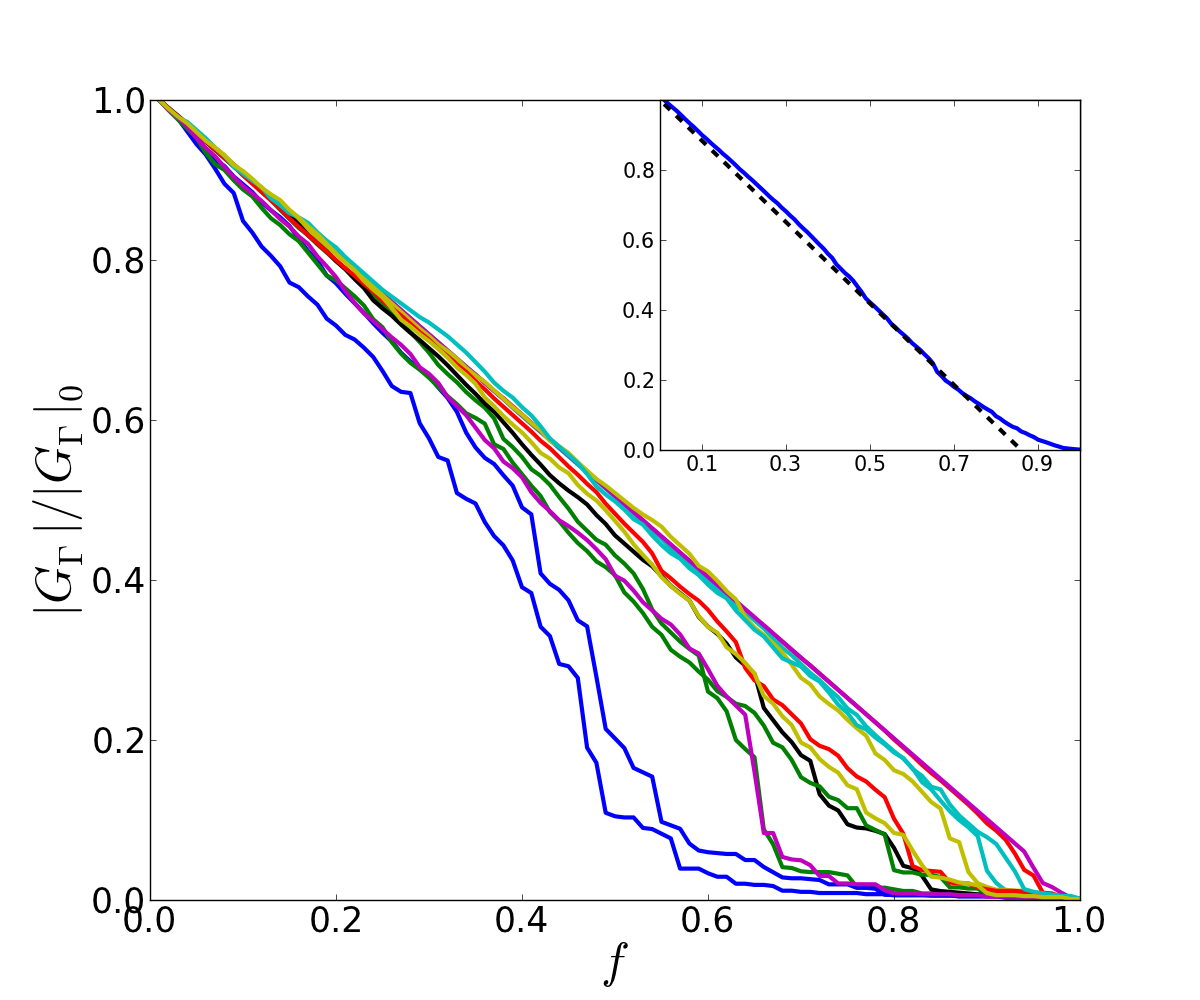}
\caption{(Color online) Sizes of the giant components of the metanetworks of evolved Boolean graphs  in genotype space vs. $f$, the fraction of nodes removed randomly,  The inset shows the average over the 16 plots.}
\label{fig:percolation}
\end{figure}

We have examined the percolation behavior of the MGE under random removal of the nodes and compared it to those of the E-R networks generated with the same edge density. For random node removal, the  generally accepted practice is that all remaining nodes at any stage  are potential targets for removal, regardless of whether they belong to the largest cluster or otherwise.  For  each successive value of the fraction $f$ of removed nodes, we have computed the largest cluster from scratch. We found that percolation behavior of the MGE (Fig.~\ref{fig:percolation}) was similar to that of the E-R network~\cite{networkpercolation,dorogovtsev2006k}.

In order to see how far these metanetworks depart from being tree-like we have calculated the clustering coefficient.
For the MGE, the clustering coefficient ranges in  $\langle C_{\rm MGE}\rangle \in ( 0.096, 0.38)$, whereas the edge density ranges in $p \in (0.002,0.044)$. 
Although MGE has a relatively high clustering coefficient, it behaves as an uncorrelated graph under random removal of nodes. This means that the percolation behavior of MGE is governed by its degree distribution. It should be noted that the clustering coefficients give information only about the immediate neighborhood of a node and it is reasonable to deduce  that these stochastic  networks have a tree-like structure at large scales as do the E-R graphs~\cite{Dorogovtsev-CCNW}. 

The dependence of the size of the giant component in the phenotype space, $|G_\Phi|$, on   $f$  is shown in Fig.~\ref{fig:percolation}, for evolved and random populations. 
 For the evolved populations at fixed $\theta=0.2$, we see that   $|G_\Phi|/|G_\Phi|_0$, decreases monotonically and essentially linearly  with $f$ (Fig.~\ref{fig:percolation}.a) and  converges to zero when $f=1$.  This linear descent to zero at $f=1$ is what we have called super robustness. 

\begin{figure}[h]
\includegraphics[width=7cm]{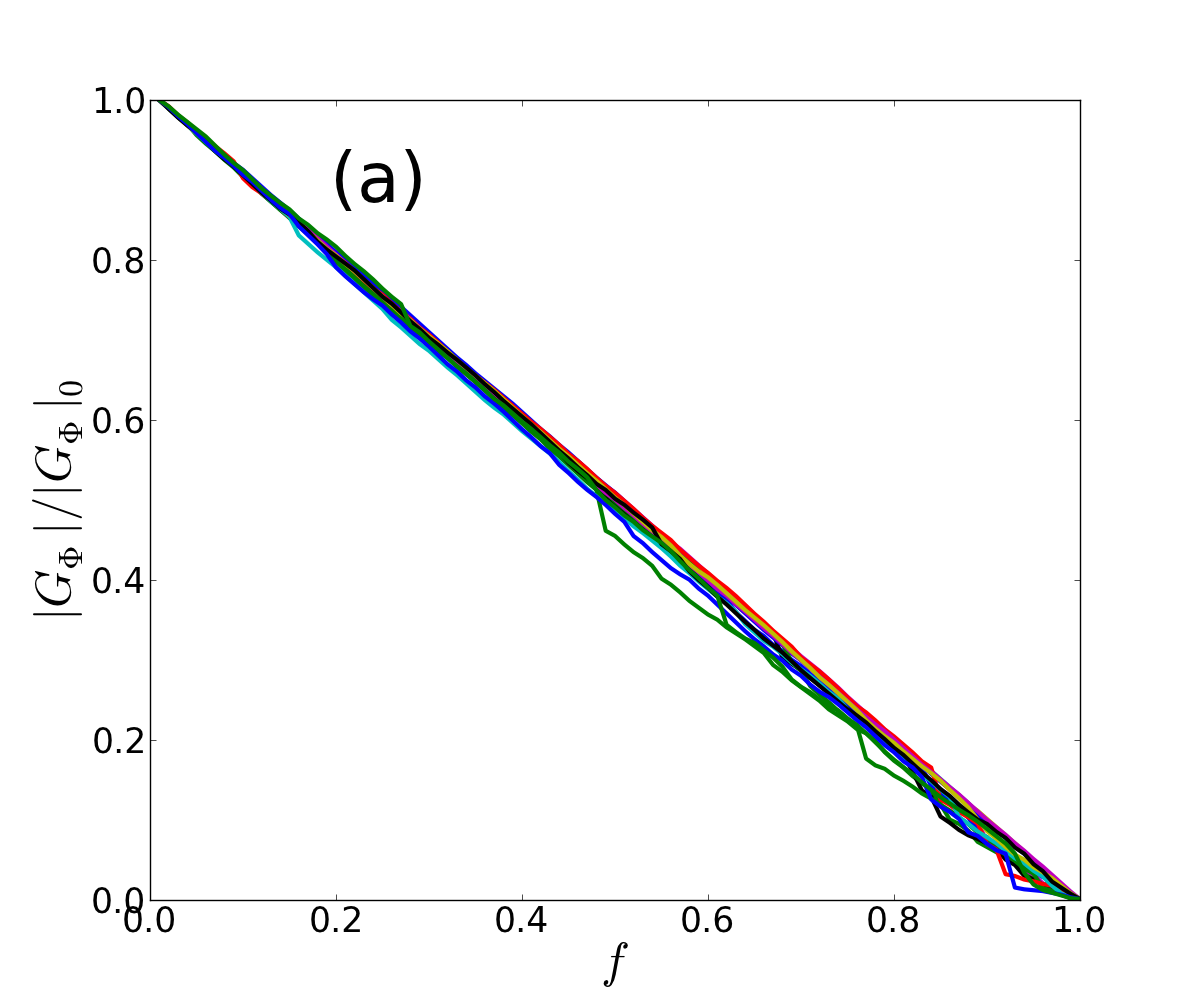}
\includegraphics[width=7cm]{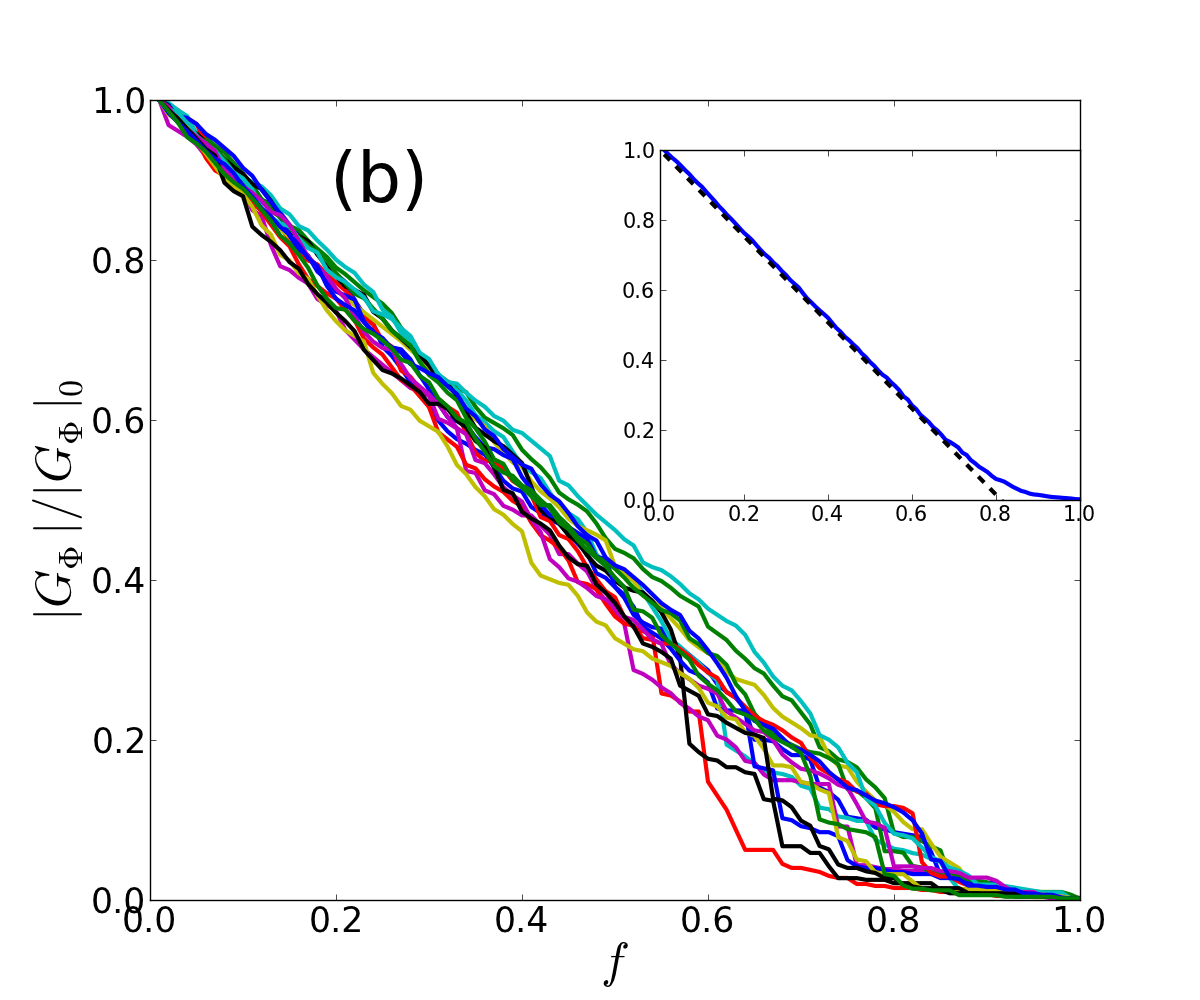}
\caption{(Color online) Sizes of the giant components of the metanetworks in phenotype space vs. $f$, the fraction of nodes removed randomly, for (a) evolved populations and (b) randomly generated graphs at $\theta= 0.2$. The inset shows the average over the 16 plots. For the evolved sets $f_c=1$, while for the random sets $f_c \simeq 0.81$, as can be seen from an extrapolation of the linear downward trend. Both plots show a linear growth of the giant component with ($f-f_c$).} 
\label{fig:percolation_MP}
\end{figure}

The MPE displays, for many individual populations,  an interesting combination of Poisson-like behaviour at very small $k$  and then a small number of  $\delta$-function peaks at some chosen $k$ values, as mentioned in Section III.  The $\delta$-peaks in the degree distribution of the individual sets correspond to cliques formed by the hubs. Under random removal of nodes, these cliques are super-resilient and do not fragment into smaller clusters. The size of the clique (the largest clique being the giant component) simply shrinks at the same rate as the total number of nodes being removed, giving rise to the linear dependence with unit negative slope that is observed in Fig.~\ref{fig:percolation_MP}.a. 

The MPR displays a percolation threshold at $f_c \simeq 0.85$ (Fig.~\ref{fig:percolation_MP}.b). For comparison we have performed surrogate simulations on E-R  graphs of the same size, and with the same  expected edge density.  These are reported in the Appendix. The E-R graph~Fig.\ref{app:ER} shows classical percolation behavior at $f_c=0.80$, in close similarity with the above figure.  

It should be noted that the nodes of the metanetwork over the populations of random graphs are not featureless and entirely interchangeable with each other. Although consisting of random networks of seven nodes each, they have a dynamics of their own, with the same Boolean keys as provided to the evolved networks.  After all, the set of attractors of the evolved and random sets are not completely disjoint, as can be seen from Fig.~\ref{fig:attractor_freqs}. Therefore we cannot totally eliminate the possibility of correlations between the edges of the MPR. 

\begin{figure}[h]
\includegraphics[width=7cm]{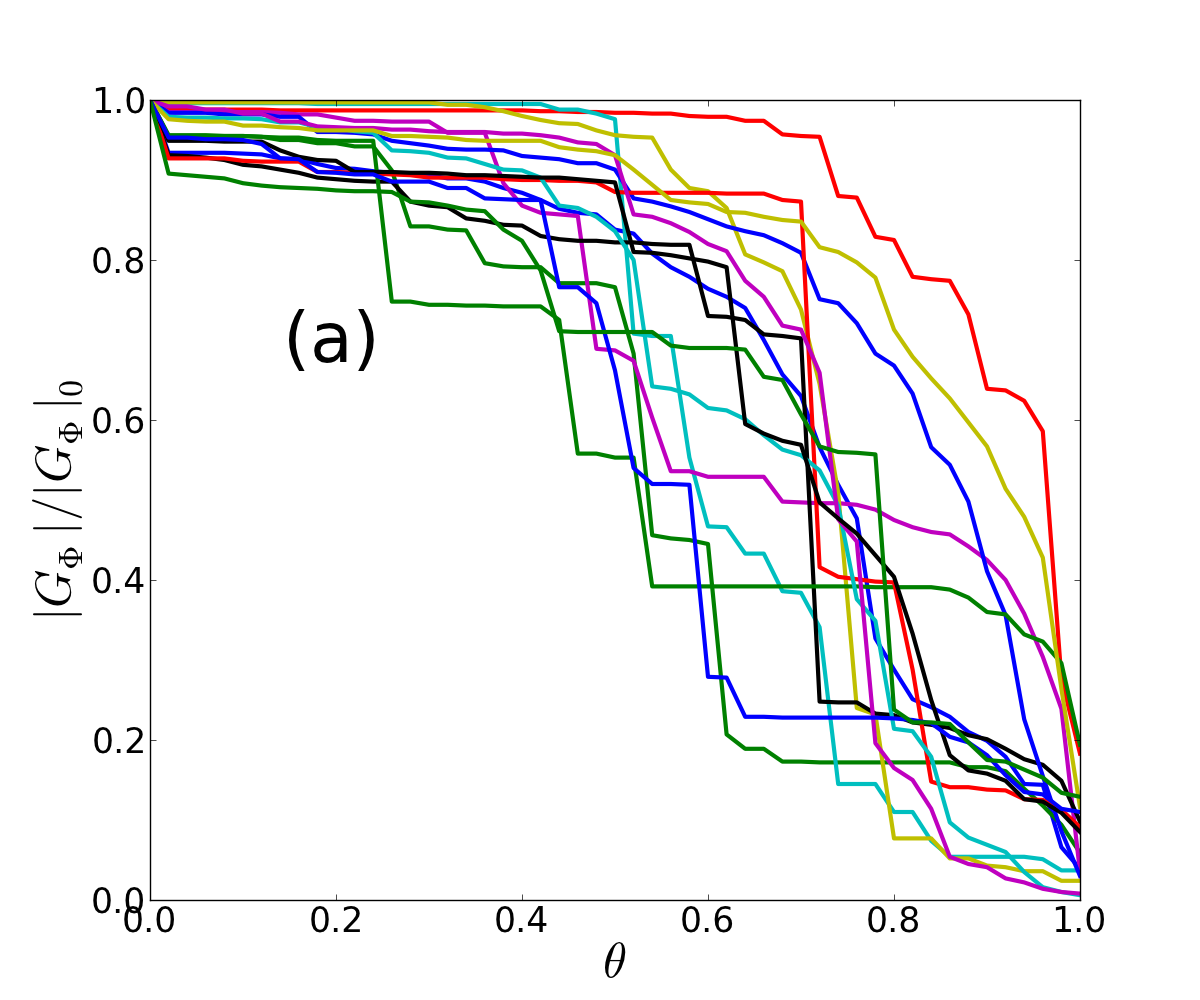}
\includegraphics[width=7cm]{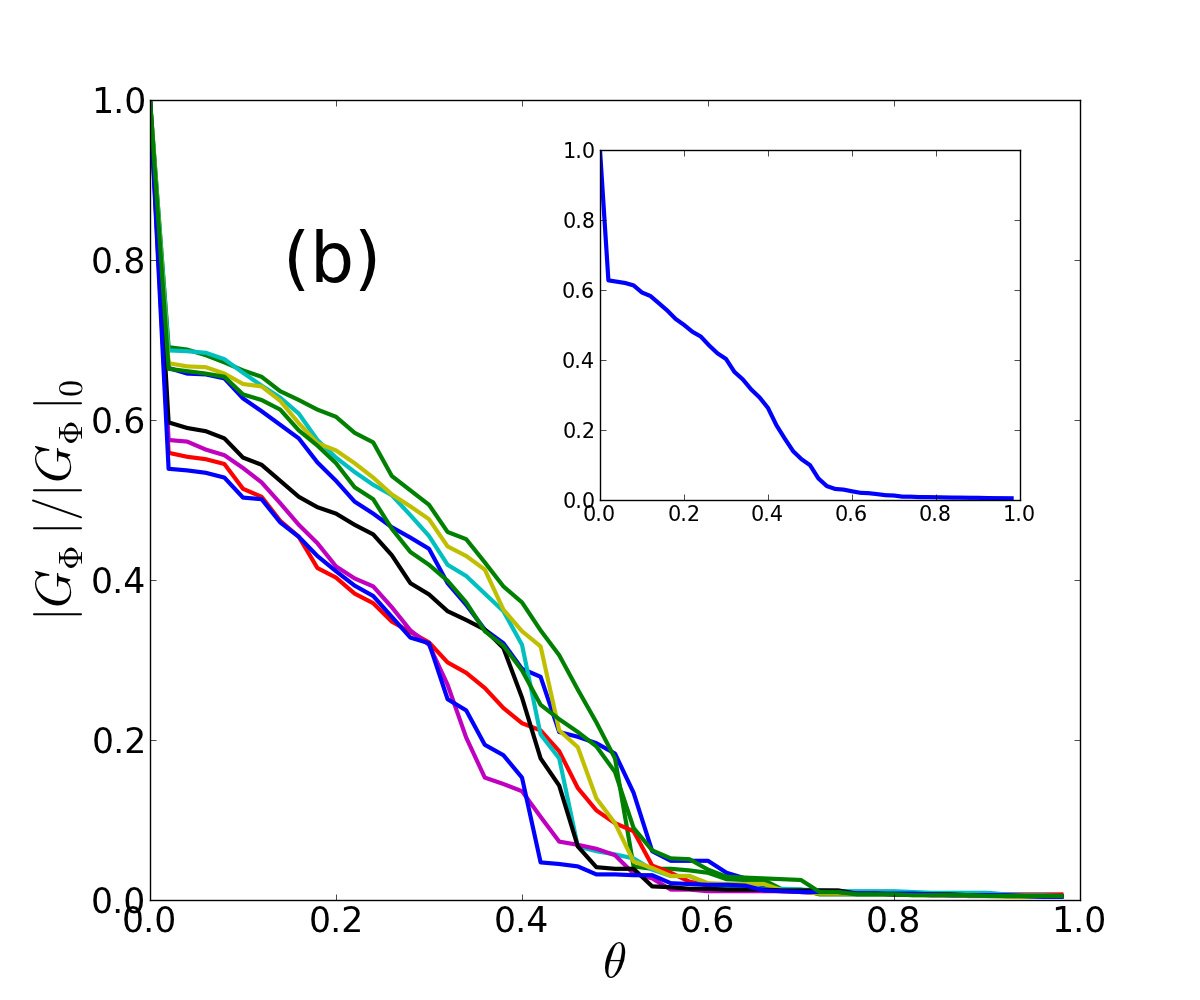}
\caption{(Color online) Sizes of the giant components of the metanetworks in phenotype space vs. threshold, $\theta$, for the evolved populations (a) and the randomly generated graphs (b).  The inset in (b) shows an average over the different random populations. The vertical axis is normalized with respect to the size of the giant component at $\theta =0$, which, in fact comprises the whole population.} 
\label{fig:giant_comps}
\end{figure}

We now examine the effect of raising $\theta$. 
The giant component of the MP comprises 100 \% of the nodes for $\theta=0$). We follow the size of this cluster, $|G_\Phi|$, as the value of $\theta$ is changed. The dependence of $|G_\Phi|$ on the threshold $\theta$ for the MPE and MPR are shown in Fig.~\ref{fig:giant_comps}, for $0 < \theta \le 1 $.  We see that the giant components of the MPRs shrink much faster than those of the MPEs, while the latter persist all the way up to $\theta=1$ and even hit the ordinate at a finite value for some sets, thanks to the  prevalence of  edges with $w=1$ (see Fig.~\ref{fig:weightdist}a). 

In Fig.~\ref{fig:giant_comps}, we also see that beyond $\theta =0.72$ the giant component of the MPE has shrunk in relative size to 50\% on the average. For larger $\theta$ values the MPE breaks up into relatively small strongly connected clusters, each of which can be thought of as a small population with a distinct phenotype. It should be stressed that changing the value of $\theta$ has no effect on the MGE. The giant component of the MGE continues to span 89\% of the total population on the average (except for the three populations out of 16 which do not have large connected components in their MGE). This means that even when the phenotype metanetwork has fragmented under very stringent conditions (large $\theta$) the giant component of the MGE connects essentially all of the different phenotypic clusters. 

The average over all the 16 sets of random populations, on the other hand, shown in the inset of Fig.~\ref{fig:giant_comps}b, exhibits a sharp critical  threshold $\theta_c \simeq 0.53$. For $\theta< \theta_c$,  the growth of the giant component is non-linear in $\theta_c-\theta$. 
%

%
\begin{table*} [!ht] 
\caption{ Numerical values of the  for 16 evolved and random populations. The column headings are, $n$, population index; ${\langle k \rangle_B}$, the mean degree of the Boolean graphs averaged over the evolved population in the steady state regime (at the $400$th step); ${\sigma_{k,B}}$, standard deviation of the degree distribution; $\langle a \rangle$, the mean attractor length averaged over all initial states as well as over the whole population; ${\langle a \rangle}_S$, $a$ averaged over a hundred steps ($300 < t < 400$) within the steady state regime; ${\sigma_a}$, standard deviation of $a$; ${\langle k \rangle_\Gamma}$ and ${\langle k \rangle_\Phi}$, the average degree of the metanetwork formed in genotype space and in phenotype space respectively, and their respective standard deviations,  ${\sigma_{k,\Gamma}}$,  ${\sigma_{k,\Phi}}$; ${\langle \kappa \rangle_\Gamma}$ and ${\langle \kappa \rangle_\Phi}$, the average shell-affiliation of the nodes of the metanetwork fo!
 rmed in genotype space and in phenotype space respectively,  and their respective standard deviations,  ${\sigma_{\kappa,\Gamma}}$,  ${\sigma_{\kappa,\Phi}}$.  Those symbols with the additional subscript $r$ pertain to relevant numerical results for random populations generated with the same edge density  as the evolved sets (therefore having the same  average degree $\langle k \rangle_B $ and standard deviation ${\sigma_{k,B}}$), those symbols without subscript $r$ pertain to evolved populations. Note that the random graphs do not form   an extensive metanetwork in genotype space. }

\begin{center}
	\begin{tabular}[c]{c  c c c c c c c c  c c c c c c c c c c}
		\hline\hline
     		 	{n}  \qquad   &   {${\langle k \rangle}_B$}   \qquad  & {${\sigma_{k,B}}$}   \quad  &   {${\langle a \rangle}_S$}  \qquad  & {${\sigma_{a_S}}$}  \quad &   {${\langle k \rangle_\Gamma}$}  \qquad  & {${\sigma_{k,\Gamma}}$}  \quad &   {${\langle k \rangle_\Phi}$}  \qquad  & {${\sigma_{k,\Phi}}$}  \quad &   {${\langle k \rangle_{\Phi,r}}$}  \qquad  & {${\sigma_{k,\Phi,r}}$}  \quad &   {${\langle \kappa \rangle_\Gamma}$}   \qquad  & {${\sigma_{\kappa,\Gamma}}$}  \quad  &   {${\langle \kappa \rangle}_\Phi$}   \qquad  & {${\sigma_{\kappa,\Phi}}$}   &  {${\langle a \rangle_r}$}  \qquad   &    {${\langle \kappa \rangle_{\Phi,r}}$}  \qquad  & {${\sigma_{\kappa,\Phi,r}}$}      \\      \\

		\hline
{ 1 } & {  2.86  }  & {  0.06  }& {  1.8  }  & {  0.04  }& {  2.2  }& {  1.02  }& {  167.69  }& {  134.16  }& {  13.0  }& {  17.3  }& {  1.92  }  & {  0.66  }& {  140.01  }  & {  105.55  }& {  3.41  } & {  9.28  }  & {  10.8  }&       \\
{ 2 } & {  2.44  }  & {  0.58  }& {  1.51  }  & {  0.03  }& {  4.05  }& {  1.98  }& {  118.39  }& {  96.08  }& {  19.54  }& {  25.96  }& {  2.96  }  & {  0.95  }& {  90.82  }  & {  62.19  }& {  3.14  } & {  12.2  }  & {  12.78  }&       \\
{ 3 } & {  3.43  }  & {  0.0  }& {  1.24  }  & {  0.02  }& {  6.22  }& {  2.65  }& {  598.39  }& {  276.63  }& {  10.26  }& {  15.02  }& {  4.18  }  & {  0.92  }& {  561.5  }  & {  260.69  }& {  3.94  } & {  7.82  }  & {  10.44  }&       \\
{ 4 } & {  2.86  }  & {  0.05  }& {  1.4  }  & {  0.01  }& {  44.6  }& {  12.66  }& {  315.76  }& {  143.8  }& {  14.61  }& {  18.92  }& {  27.7  }  & {  3.23  }& {  247.5  }  & {  99.61  }& {  3.35  } & {  10.49  }  & {  12.03  }&       \\
{ 5 } & {  3.29  }  & {  0.0  }& {  1.47  }  & {  0.02  }& {  17.11  }& {  5.36  }& {  422.02  }& {  254.15  }& {  9.9  }& {  14.71  }& {  10.43  }  & {  1.15  }& {  359.49  }  & {  208.08  }& {  3.91  } & {  7.35  }  & {  9.8  }&       \\
{ 6 } & {  2.57  }  & {  0.0  }& {  1.83  }  & {  0.02  }& {  69.6  }& {  18.12  }& {  282.57  }& {  127.12  }& {  15.37  }& {  21.05  }& {  44.67  }  & {  5.42  }& {  237.64  }  & {  105.39  }& {  3.24  } & {  10.41  }  & {  12.1  }&       \\
{ 7 } & {  3.02  }  & {  0.04  }& {  1.73  }  & {  0.04  }& {  2.53  }& {  1.23  }& {  137.23  }& {  120.08  }& {  10.84  }& {  16.21  }& {  2.09  }  & {  0.78  }& {  127.18  }  & {  112.3  }& {  3.65  } & {  7.78  }  & {  10.23  }&       \\
{ 8 } & {  3.84  }  & {  0.05  }& {  1.78  }  & {  0.02  }& {  11.58  }& {  5.73  }& {  615.74  }& {  250.64  }& {  10.48  }& {  16.62  }& {  7.3  }  & {  2.52  }& {  542.65  }  & {  208.6  }& {  4.3  } & {  8.15  }  & {  11.7  }&       \\
{ 9 } & {  2.86  }  & {  0.0  }& {  1.54  }  & {  0.03  }& {  7.1  }& {  2.81  }& {  243.78  }& {  158.77  }& {  14.09  }& {  18.96  }& {  4.61  }  & {  0.91  }& {  221.36  }  & {  145.16  }& {  3.37  } & {  9.9  }  & {  11.7  }&       \\
{ 10 } & {  3.14  }  & {  0.0  }& {  1.64  }  & {  0.04  }& {  4.76  }& {  2.33  }& {  147.02  }& {  68.45  }& {  11.37  }& {  16.11  }& {  3.33  }  & {  1.02  }& {  128.99  }  & {  51.8  }& {  3.65  } & {  8.43  }  & {  10.9  }&       \\
{ 11 } & {  2.76  }  & {  0.32  }& {  1.73  }  & {  0.03  }& {  5.67  }& {  3.79  }& {  299.78  }& {  214.52  }& {  15.24  }& {  20.56  }& {  3.89  }  & {  1.77  }& {  256.73  }  & {  188.09  }& {  3.41  } & {  10.44  }  & {  11.8  }&       \\
{ 12 } & {  2.72  }  & {  0.04  }& {  1.68  }  & {  0.04  }& {  9.32  }& {  4.26  }& {  375.19  }& {  161.88  }& {  14.09  }& {  19.3  }& {  5.94  }  & {  1.56  }& {  249.58  }  & {  72.92  }& {  3.5  } & {  9.87  }  & {  11.61  }&       \\
{ 13 } & {  3.4  }  & {  0.46  }& {  1.51  }  & {  0.03  }& {  2.55  }& {  1.3  }& {  487.3  }& {  275.52  }& {  10.31  }& {  15.72  }& {  2.13  }  & {  0.84  }& {  451.12  }  & {  252.02  }& {  4.02  } & {  7.56  }  & {  10.33  }&       \\
{ 14 } & {  2.99  }  & {  0.2  }& {  1.7  }  & {  0.03  }& {  18.41  }& {  15.25  }& {  401.11  }& {  226.85  }& {  11.97  }& {  15.92  }& {  11.57  }  & {  8.2  }& {  414.41  }  & {  251.23  }& {  3.47  } & {  8.67  }  & {  10.19  }&       \\
{ 15 } & {  2.29  }  & {  0.0  }& {  1.71  }  & {  0.02  }& {  5.57  }& {  2.51  }& {  119.94  }& {  87.26  }& {  19.56  }& {  25.49  }& {  3.83  }  & {  0.99  }& {  100.54  }  & {  77.41  }& {  2.94  } & {  12.98  }  & {  14.5  }&       \\
{ 16 } & {  3.0  }  & {  0.03  }& {  1.85  }  & {  0.03  }& {  6.78  }& {  3.7  }& {  152.48  }& {  154.05  }& {  14.51  }& {  19.56  }& {  4.48  }  & {  1.63  }& {  140.74  }  & {  147.31  }& {  3.48  } & {  10.53  }  & {  12.57  }&       \\

			\hline\hline
		\end{tabular}
	\end{center}
\label{tab:table}
\end{table*}

\section{Conclusion}
Nimwegen {\it et al.}~\cite{Nimwegen} have argued that under neutral evolution modeled by different types of random walks, the population  will tend to concentrate in regions of the genotype space with  high mutational robustness (also see \cite{Wagner2}), defined as  the average degree of the neutral network weighted by the size of the  populations residing on the nodes.  Clearly this corresponds to an increase  in the average degree, in the course of evolution.  
This is indeed what we find in the model system which we have presented here.

Random Boolean graphs  (BG) do not form a  connected cluster spanning any appreciable portion of the  genotype space and therefore cannot explore the phenotype space effectively with single-mutation steps. 
Under the genetic algorithm selecting for short attractors, the population of random BG evolves in such a way that the giant connected cluster spans 60-100\% of all the genotypes for almost all the populations, which means that it also spans almost all of the phenotype space, making  phenotypic innovations possible. 

In our model, not just a single phenotype, but different behaviors under different conditions are possible.
We have thus  been able to provide an example of  an evolving model population which  spontaneously engenders a complex  metanetwork in phenotype as well as in  genotype space and allows the plasticity of the phenotype.  

We have found that both the MGE and the MPE maintain a high degree of connectivity in the face of both correlated removal of weak edges and random removal of nodes.    Both these processes have their analogues in evolutionary scenarios.
Filtering out phenotypic bonds weaker than a given  threshold would correspond, {\it e.g.}, to increased selection pressure sharpening the peaks in the fitness landscape.  Tightening the minimum requirements (minimum size of the basins of attraction of the shared attractors) for the identification of shared features,  corresponds to stipulating  that these features themselves should not vary due to variations in the ``initial conditions'' or environmental inputs. 

The random removal of a large fraction of nodes is similar to a large scale catastrophe which indiscriminately destroys most of the population. 
The MPE is ``super robust'' under random elimination of nodes, with the relative size of the giant component tending only  linearly  (rather than exponentially) to zero as the fraction of removed nodes approaches unity. Thus, in an evolved population, even for a death rate approaching unity, a finite fraction of the largest connected cluster in phenotype space will survive, besides isolated individuals. For a random population, all the survivors, if any, will be  phenotypically disconnected.  The former case means that, in a  ``great catastrophe'' scenario,  a genetically related and phenotypically similar  small community  has a finite chance for survival. 

We would like to remark here  that robustness, understood as the mean degree of the neutral network~\cite{Kimura,Nimwegen} is a surpisingly successful mean-field approach to the problem of evolvability.  Although the MGE has a higher than expected clustering coefficient for an essentially random, E-R like graph, this is a relatively short-range effect, and at larger scales it has a tree-like structure~\cite{Dorogovtsev-CCNW}.  This is evident from a comparison between the response to random node removal of the MGE in Fig.~\ref{fig:percolation} and of the surrogate E-R graph, Fig.~\ref{app:ER}, both displaying a linear decay with $f$, modified only close to the critical threshold by weak or strong clustering properties.~\cite{networkpercolation,dorogovtsev2006k,networkpercolation}

It is known that tree-like networks are the  most  efficient structures for spanning a given set of nodes, and are the basis of efficient search strategies on an unknown network.  The efficiency goes up with the Cheeger constant~\cite{Cheeger}, defined as the minimum ratio of the ``surface'' nodes of a non-trivial subset of the network to the number of nodes contained within such subsets; it is thus a measure of the ``expansion'' of the graph.  Thus, a MGE which may have appreciable clustering at short scales but a Poisson degree distribution and a  tree-like structure at large scales  makes it very efficient in probing the phenotype space.  

The  degree and strength distributions of the MPE vary with the chosen  weight threshold $\theta$.
In Section IVA, one sees that in the ensemble of similarly generated populations, there is  a regularity which is missing from any given population. The degree distribution of the {\it combined} populations shows that, for small values of  $\theta$ (say $\theta=0.2$), the averaged probability of having a degree $k$, {\it increases} with increasing $k$, having a putative power law scaling form $p_k \sim k^{-\gamma}$, with  $ \gamma < 0$.  This   gives rise to a predominance of hubs.  
For larger $\theta$, the exponent gradually crosses over to $\gamma>0$ (Fig.~\ref{fig:alldegdist_allthetas}), although the first moment is still divergent. 

{\bf Acknowledgements}

We are grateful to Reka Albert and Alain Barrat for some  useful exchanges.  
\vskip 0.5cm

{\bf Appendix}

\appendix
\setcounter{figure}{0} \renewcommand{\thefigure}{A.\arabic{figure}}
\setcounter{table}{0} \renewcommand{\thetable}{A.\arabic{table}}
\subsection{\bf Finite size effects}
To estimate the finite size effects in the percolation behavior for random removal of vertices in the MPR, we have done independent simulations over 16 different sets of  Erd\"os Renyi (E-R) graphs  with the same number ($10^3$) of  nodes and edge density  ($\langle k \rangle/N= 0.0046$) as in Fig.~\ref{fig:percolation}b. 
In Fig. A.1, the percolation threshold can be read off from extrapolating the linear part of the average curve in the inset, to be $f_c=0.80$. This matches  the value which we obtain by following the steepest slopes of the individual curves and is also very close to the percolation threshold found  in Fig.~\ref{fig:percolation_MP}b.

\subsection{\bf Definitions: Dynamical systems and Networks}
A finite dynamical system with discrete states  is called an {\it automaton}.  An automaton may be represented as a network consisting of nodes and edges.  The nodes may take on different values.  In this paper the nodes of the gene regulatory networks (GRNs) will be modeled by the discrete, Boolean values  0 or 1, corresponding to a gene being off or on. As explained in Section II, the {\bf  state} of an automaton is a list of the values of its nodes. The edges and some logic functions at the nodes,  represent the interactions between the nodes of this {\it Boolean network}. 

\begin{figure}[ht]
\centering
\includegraphics[width=7.5cm]{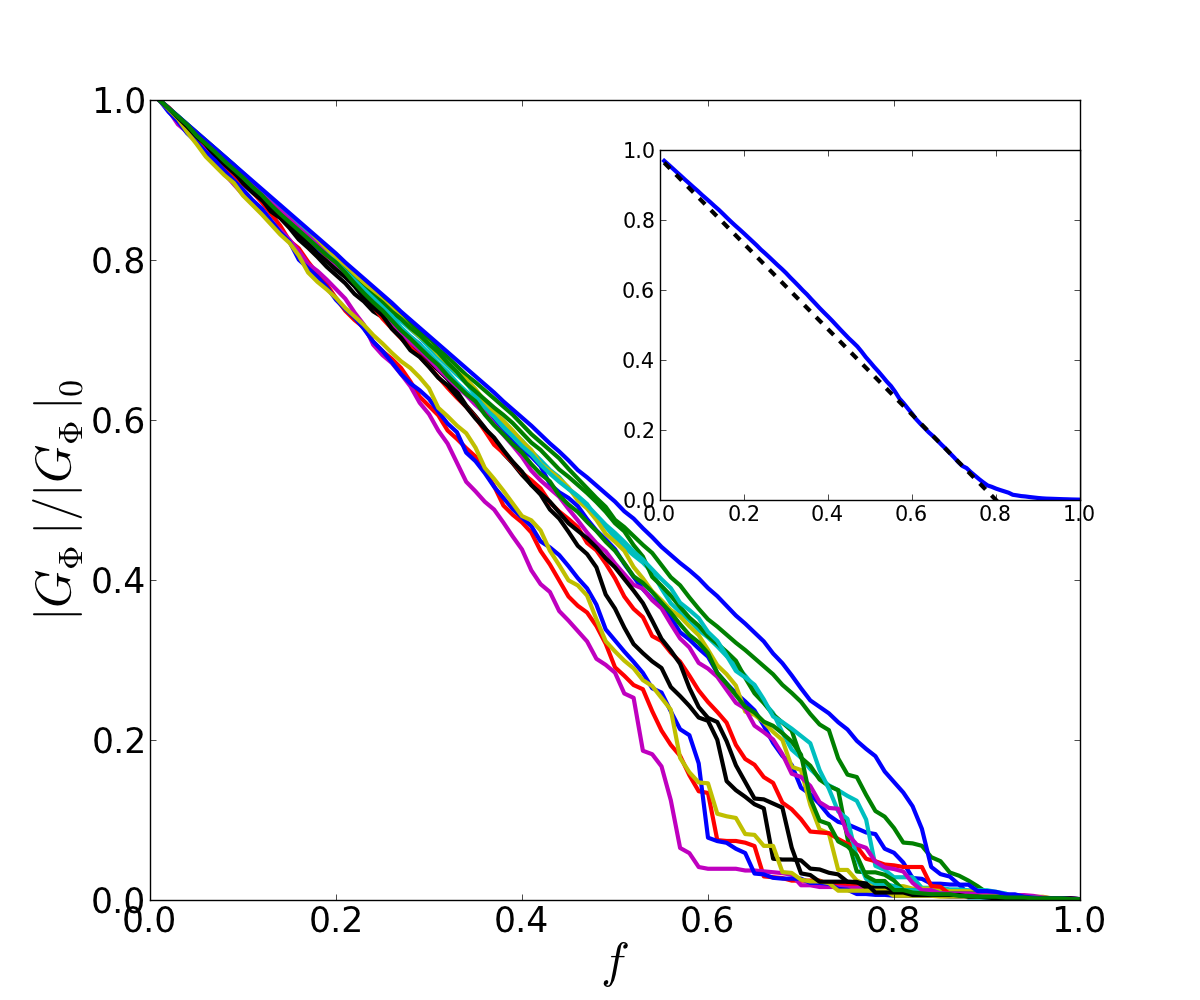}
\caption{Percolation behaviour of 16 independently generated random (Erd\"os-Renyi) graphs  the same size  and  initial edge density as for the random Boolean graphs in Fig.~\ref{fig:percolation}b.  The inset shows an average over the 16 sets. } 
\label{app:ER}
\end{figure}
 In this paper we adopt a {\bf synchronous dynamics}, which means that given any state of the automaton, an {\bf updating rule} (see Section II) which updates the values of all the nodes simultaneously, determines  the next state of the automaton, and then the next state and so on.  The succession of states under this dynamics is called a {\bf  trajectory}. The number of possible states of an automaton is called its {\bf phase space}.  The phase space  of a such a discrete, finite automaton has to be finite (in fact $s^N$ for a graph with $N$ nodes and $s$ possible states for each node).  Therefore,  any trajectory eventually has to repeat itself, i.e.,  is at most  periodic. 

In many cases trajectories,  trajectories  starting from many different  states converge on one {\bf fixed point} where  all change ceases.  This state is called a {\bf point attractor}. There may also be  collections of states  whose trajectories  eventually end up  on an ordered set of  states which keep cycling.  This set of states is called a {\bf periodic attractor}.  We will call the cardinality of this ordered set the {\bf length of the periodic attractor}. Clearly the length of a periodic attractor can at most be the size of the phase space, but in general it is much smaller.   A collection of states which end up either in a point or a periodic attractor is called the {\bf basin of attraction} of this attractor.   

The phase space is partitioned into as many  different basins of attraction as there are attractors. If the system is in a {\bf steady state} (fixed point or periodic attractor) and an external intervention alters  the state of the system and takes it  to some other state within the basin of attraction of this attractor, then the system will return to its steady state.  It may be that the intervention takes the system to a state in a different basin of attraction.  In this case it will flow to a new attractor.    There may also exist isolated fixed points, to which no trajectory flows; these are not  {\bf stable}, i.e., once perturbed to any neighboring state, the system will eventually find itself in some other fixed point or periodic attractor. 

A {\bf network} consists of {\bf nodes} and {\bf edges}, which may be directed or undirected, connecting the nodes.  A network can alternatively be called a {\bf graph}. In this paper we study {\bf metanetworks}, which are networks of networks.  To avoid   confusion we have used the term {\bf graphs} for the small networks which constitute the nodes of the metanetwork. 

	 The number of edges connecting a node to other nodes  (or to itself) is called its {\bf degree}.  The  {\bf degree distribution}, denoted by $p_k(k)$ in this article, is the probability of encountering a node with the degree $k$. The mean degree is $ \langle k \rangle =\sum_{i=1, \ldots, N} k p_k (k)$.   The {\bf clustering coefficient} of a node with $k$ neighbors  is defined as  $C_k = 2e/[k(k-1)]$; $e$ is the number of edges interconnecting  its neighbors and it is normalized by the number of distinct pairs of neighbors.  The {\bf average clustering coefficient} of a network is defined as $C=\sum_k p(k) C_k$. 

It is possible to  decompose a network into successive shells of lower to higher connectivity.  This is termed {\bf $k$-core analysis}.~\cite{Bollobas1998,Batagelj2003}   We have used the Greek letter $\kappa$ in this article so that there is no confusion with the degree $k$. The prescription is  as follows:  {\it i)} Disconnect all nodes which have degree one.  {\it ii)} Repeat the process until no nodes of degree one remain. {\it iii)} Label all these nodes as the {\bf 1st shell}.  {\it iv)} Repeat this process for degree 2, 3, etc., labeling the severed nodes accordingly as belonging to the $\kappa = 2,3,\ldots$-shell  until no nodes remain. 

In network theory,  a network is termed {\bf robust} under random removal or nodes (edges), if it retains a connected component containing most of the nodes even in very late stages of decimation, the proportion  of removed nodes (edges) before total breakdown tending to unity as the number of nodes tends to infinity.  This behavior has been demonstrated for scale free networks by Albert and Barabasi~\cite{Albert}. Random networks, in contrast, disintegrate into many nodes or very small clusters at intermediate stages of node (edge) removal.  
%

\bibliographystyle{apsrev4-1}
\bibliography{Danaci_metanetworks_v2}

\end{document}